# Morphoelastic ribbons: Differential growth-induced curvature and torsion

Hao Liu [a], Mingwu Li [b], Dabiao Liu [a, c, *]

[a] *Department of Engineering Mechanics, School of Aerospace Engineering, Huazhong University of Science and Technology, Wuhan 430074, China*

[b] *Department of Mechanics and Aerospace Engineering, Southern University of Science and Technology, Shenzhen 518055, China*

[c] *Hubei Key Laboratory of Engineering Structural Analysis and Safety Assessment, Wuhan 430074, China*

* Corresponding author: Dabiao Liu
E-mail address: dbliu@hust.edu.cn

## Abstract

Natural slender structures, such as plant leaves, petals, and tendrils, often exhibit complex three-dimensional (3D) morphologies—including twisting, helical coiling, and saddle-bending—driven by differential growth. The resulting internal stresses are partially relieved through the development of intrinsic curvature and torsion. The fundamental challenge lies in effectively correlating microscopic growth fields to the macroscopic shapes and mechanical responses of the ribbon structures. However, existing ribbon or shell models struggle to directly link growth gradients to macroscopic curvature and torsion, necessitating a reduced-dimensional framework. This work establishes a unified one-dimensional (1D) morphoelastic ribbon model derived rigorously from 3D finite elasticity theory via a two-step asymptotic dimension reduction. The reduced-order model captures key geometric nonlinearities and finite rotations while retaining explicit dependence on the growth tensor. We obtain analytical solutions for saddle-bending and twisting configurations induced by specific growth gradients. Furthermore, numerical continuation, based on the reduced model, reveals post-buckling transitions into helical morphologies, identifying bifurcation thresholds and constructing phase diagrams. This framework explicitly links growth fields to ribbon curvature and torsion, providing fundamental mechanics insights into the morphogenesis of slender plant organs and offering the potential for bioinspired soft robotics design.

# 1 Introduction

Plant leaves and petals exhibit remarkable three-dimensional (3D) morphological diversity, driven by differential growth. For instance, twining plants develop helical structures through asymmetric growth rates across their cross-sections (Silk, 1989a, b). This growth-induced shape transformation underscores the need for mechanical models that link microscopic growth fields to macroscopic morphologies. The narrow petals of *Dendrobium helix* exhibit a twisting shape (see Fig. 1(a)), while those of other *Orchidaceae* species form a distinct helical configuration (see Fig. 1(b)). Recent genetic studies indicate that this type of helical growth may be associated with the helical arrangement of cortical microtubules and the overlying cellulose microfibrils (Ishida et al., 2007; Smyth, 2016; Wada, 2012). The petals of *Pancratium maritimum* bend into a saddle-bending shape (see Fig. 1(c)), possibly influenced by the differential expansion of parenchyma cells in response to environmental stimuli such as temperature, light, or humidity (van Doorn and Kamdee, 2014). Continuum mechanics principles reveal that such morphologies emerge through stress redistribution mechanisms driven by differential growth between tissues (Huang et al., 2018). This leads to internal strain mismatches and consequently induces curvature and torsion.

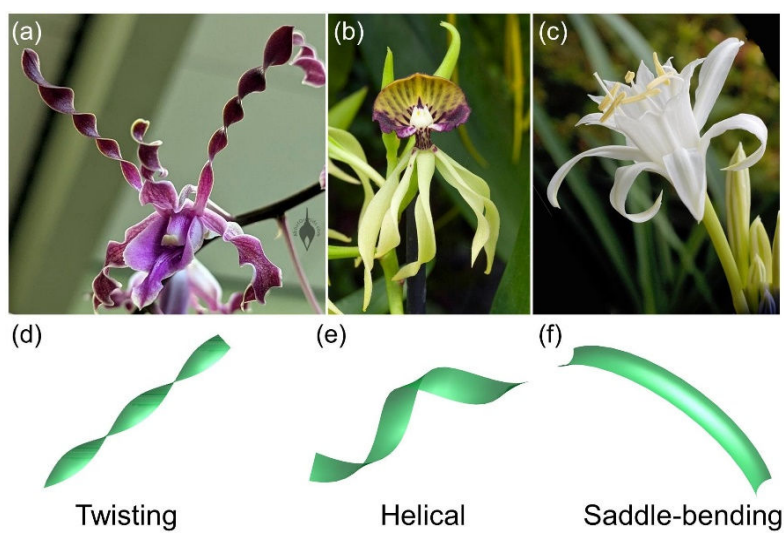

*Fig. 1. Three-dimensional morphologies. (a) Dendrobium Helix, image from <http://www.aboutorchids.com/blog/page/8/> (b) Orchidaceae, image from <https://www.pinterest.com/pin/245586985924699326/>. (c) Pancratium maritimum, image from <https://www.floralimages.co.uk/page.php?taxon=pancratium_maritimum,1>. (d) Twisting configuration. (e) Helical configuration. (f) Saddle-bending configuration.*

Plant growth is a complex, multi-scale, dynamic process. It can be considered a slow, quasi-static process that requires time to relax into its equilibrium shape (Hsu, 1968). Rodriguez et al. (1994) (Rodriguez et al., 1994) developed a general three-dimensional continuum framework for soft elastic tissues with finite volume growth. They decomposed the total deformation into growth deformation in tissue stress-free reference state and elastic deformation. Building upon this morphoelastic theory, numerous models have been developed to study the mechanical behavior of plants and biological tissues with differential growth, ranging from Föppl–von Kármán type plate/shell models (Dervaux and Ben Amar, 2008; Dervaux et al., 2009; Wang et al., 2024; Yin et al., 2021) to asymptotically consistent morphoelastic shells (Wang et al., 2018; Yu and Chen, 2024). However, the complexity of these plate/shell models makes it difficult to intuitively relate the local growth fields of slender structures to their macroscopic curvatures, torsion, and elongation. For slender structures, such as petals, plant leaves, and stems, where the characteristic scale is significantly smaller than the length, one-dimensional (1D) models offer a more suitable and efficient approach.

Due to the particular aspect ratio of slender structures, their deformations can be described through the centerline using differential geometry and balance laws. For instance, the classical Kirchhoff rod model (Dill, 1992; Kirchhoff, 1859), based on the assumption of linear small strains, incorporates the curvatures along the centerline into the strain energy of rods. The generalized Cosserat rod model simultaneously accounts for axial extension and shear deformation of the centerline (Cosserat and Cosserat, 1909). These rod models have been successfully applied across a wide range of scales, from the motion of DNA (Wolgemuth et al., 2000; Yang et al., 1993) to buckled filaments (Charles et al., 2019; Liu et al., 2024).

To connect growth fields with the deformation of slender structures, Moulton et al. (2013) formulated the problem under the assumption of purely axial growth, prescribing intrinsic curvatures, and recast it as a static equilibrium problem governed by the Kirchhoff equations with linear constitutive laws. These equations have been applied to study the waving, coiling, and skewing patterns of Arabidopsis thaliana roots on inclined surfaces with appropriate boundary conditions (Porat et al., 2024; Sipos and Várkonyi, 2022). Later, Moulton et al. (2020) modeled the rods as hyperelastic morphoelastic solids capable of growth, remodeling, and supporting cross-sectional deformations. By introducing the small aspect-ratio parameter for asymptotic expansion, they solved the cross-sectional deformations via energy minimization, resulting in a 1D energy expression analogous to that of an elastic rod. Unlike the Kirchhoff model, the present formulation incorporates fully geometrically nonlinear strain measures. The resulting models significantly simplify the analysis while preserving key mechanical characteristics of the original system, including multistability and the associated bifurcation behavior. The key step in deriving the model is the dimension reduction process, which systematically utilizes the slenderness of structures in one or more directions to derive lower-dimensional models.

Dimension reduction has been applied to various mechanical systems, such as tensile necking in prismatic solids (Audoly and Hutchinson, 2016; Coleman and Newman, 1988; Fu et al., 2021), bulging of hyperelastic tubes caused by internal pressure (Lestringant and Audoly, 2018; Ye et al., 2020; Yu and Fu, 2023), and the active filaments in elastic rods (Kaczmarski et al., 2024; Kaczmarski et al., 2022). Audoly and Lestringant (2021) proposed a relaxation approach for deriving an asymptotically exact 1D model for slender structures in a fully nonlinear setting. This method was applied to deriving 1D models for ribbons (Audoly and Neukirch, 2021; Gomez et al., 2023) and tape springs (Kumar et al., 2023). Unlike elastic rods, ribbons are distinguished by their cross-sectional geometry: their thickness is much smaller than their width, which is, in turn, much smaller than their length. As a result, ribbons can exhibit large global deformations similar to those of rods, while also displaying distinct features such as isometric deformations and stress localization (Audoly and Van der Heijden, 2023; Levin et al., 2021; Pham Dinh et al., 2016). Extensive experimental and theoretical studies have been performed to study the nonlinear torsional behaviors and morphological evolutions of ribbons, mostly focusing on equilibrium configurations under prescribed boundary conditions, without incorporating active effects such as growth (Chopin and Kudrolli, 2013, 2022; Dai et al., 2025; Liu et al., 2025; Liu et al., 2022; Yang et al., 2025; Yu and Hanna, 2019). However, a unified framework for describing the spatial configurations and equilibrium states of ribbons under differential growth fields remains lacking.

This paper aims to develop a 1D morphoelastic ribbon model by combining morphoelastic theory with dimension reduction, starting from 3D elasticity. The model links growth fields to macroscopic curvature and torsion, while asymptotically capturing nonlinear strain and finite rotation. The rod-like formulation supports Kirchhoff rod equations and enables an efficient numerical implementation, offering a fast prediction of ribbon morphology while preserving local cross-sectional details.

The rest of this paper is arranged as follows. In Section 2, by rigorously applying nonlinear elasticity theory and the dimension reduction framework, we systematically reduce the 3D governing equations to a 1D ribbon model with differential growth. This reduced-order model asymptotically captures internal strains and finite rotations induced by growth. Analytical solutions of growth-induced saddle bending and twisting configuration are then deduced in Section 3. Numerical simulations are also implemented to capture the evolution of curvature and torsion in helical configurations induced by growth strain. Conclusions and limitations are drawn in Section 4. Throughout this paper, the summation convention for repeated indices is adopted. Specifically, Greek indices such as $\alpha$ and $\beta$ run from 2 to 3, while Latin indices including $i$ and $j$ range from 1 to 3. A comma preceding an index denotes differentiation.

## 2 Morphoelastic ribbon

In this Section, we derive the 1D morphoelastic ribbon model from 3D nonlinear elasticity through a two-step asymptotic analysis. We begin by recalling the geometric formulation of a ribbon based on a centerline representation. Following the multiplicative decomposition proposed by Rodriguez et al. (1994) for growth-induced large deformations, we formulate the strain energy of the hyperelastic ribbon. By introducing two small parameters—the thickness-to-width ratio and the slenderness ratio—we systematically reduce the 3D model to a 2D shell and then to a 1D ribbon model. The variational formulation of the energy provides explicit expressions for the microscopic displacements, ultimately yielding the equilibrium equations and boundary conditions for the morphoelastic ribbon.

## 2.1 Kinematics of the ribbon

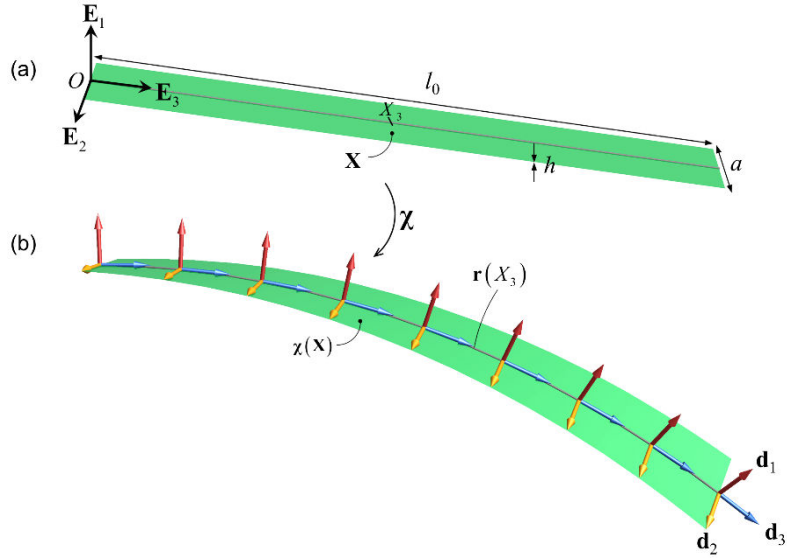

*Fig. 2 Kinematic description for the ribbon based on the centerline. (a) Reference configuration. (b) Deformed configuration.*

We consider an initial flat ribbon configuration $\mathcal{B}_0 \in \mathbb{R}^3$ that occupies a region $\Omega_0 \times [-h/2, h/2]$ in its reference configuration, where $\Omega_0$ is the middle surface of the ribbon; see Fig. 2 (a). A Cartesian coordinate system is established so that the position of an arbitrary point can be described by $\mathbf{X} = X_i \mathbf{E}_i$. Here, $\mathbf{E}_i$ is the basis vector. The origin $O$ is located at the center of the left end, and $\mathbf{X} = (X_1, X_2, X_3)$ is the coordinate along the thickness, width, and length directions, respectively. The dimensions of the ribbon are length $l_0$, width $a$, and thickness $h$ in the reference configuration.

The deformations of the ribbon in the current configuration $\mathcal{B}$ is described by $\mathbf{x} = \boldsymbol{\chi}(\mathbf{X})$. The mapping $\boldsymbol{\chi}: \mathcal{B}_0 \to \mathcal{B}$ can be expressed by the centerline $\mathbf{r}(X_3)$ associated with a set of orthogonal directors; see Fig. 2 (b). From this centerline, we define a local orthogonal director basis $(\mathbf{d}_1(X_3), \mathbf{d}_2(X_3), \mathbf{d}_3(X_3))$, where the director $\mathbf{d}_3(X_3)$ is chosen to be the unit tangent of the centerline

$$\mathbf{d}_3(X_3) = \frac{\mathbf{r}'(X_3)}{|\mathbf{r}'(X_3)|}. \tag{1}$$

Here and through this paper, we use primes $(\cdot)'$ to denote the derivatives with respect to $X_3$. And $\varepsilon = |\mathbf{r}'(X_3)| - 1$ is the axial strain of the centerline. The orthonormality of the directors implies that

$$\mathbf{d}_i(X_3) \cdot \mathbf{d}_j(X_3) = \delta_{ij}, \tag{2}$$

where $\delta_{ij}$ is Kronecker's symbol.

The centerline $X_3 \mapsto \mathbf{r}(X_3)$ of the ribbon passing through the centroid of each one of cross-section in the current configuration makes the condition

$$\mathbf{r}(X_3) = \langle \boldsymbol{\chi}(\mathbf{X}) \rangle. \tag{3}$$

Here and through this paper, the symbol $\langle f(\mathbf{X}) \rangle = \frac{1}{ah} \int_{-h/2}^{h/2} \int_{-a/2}^{a/2} f \, dX_1 dX_2$ extracts the average of a function $f$ over the cross-section for each $X_3$. The orientation of $\mathbf{d}_1(X_3)$ and $\mathbf{d}_2(X_3)$ in the plane perpendicular to $\mathbf{d}_3(X_3)$ is fixed by the condition (Audoly and Lestringant, 2021)

$$\forall X_3 \in [0, l_0]: \left\langle \left(X_1 \mathbf{d}_1(X_3) + X_2 \mathbf{d}_2(X_3)\right) \times \left(\boldsymbol{\chi}(\mathbf{X}) - \mathbf{r}(X_3)\right) \right\rangle \cdot \mathbf{d}_3(X_3) = 0. \tag{4}$$

From the director basis, we define the Darboux vector $\boldsymbol{\omega}(X_3) = \kappa_i(X_3) \mathbf{d}_i(X_3)$. This vector describes the rotation gradient of the director frame with respect to the centerline, satisfying

$$\mathbf{d}_i'(X_3) = \boldsymbol{\omega}(X_3) \times \mathbf{d}_i(X_3). \tag{5}$$

Here, the quantities $\kappa_1(X_3)$ and $\kappa_2(X_3)$ are the bending curvatures, $\kappa_3(X_3)$ is the twisting curvature.

## 2.2 Elastic strain energy with growth

By differentiating the deformation $\chi(\mathbf{X})$ with respect to $\mathbf{X}$, we have the deformation gradient $\mathbf{F} = F_{ij}\mathbf{d}_i \otimes \mathbf{E}_j$, where the symbol "$\otimes$" denotes the tensor product. Based on multiplicative decomposition proposed by Rodriguez et al. (1994) for growth-induced large deformation, the deformation gradient tensor is decomposed as

$$\mathbf{F} = \mathbf{A} \cdot \mathbf{G}, \tag{6}$$

where $\mathbf{G}$ is the growth tensor describing the addition or removal of material mass to the body; and $\mathbf{A} = \mathbf{F} \cdot \mathbf{G}^{-1}$ is the elastic deformation tensor ensuring the compatibility of the body. We further decompose the growth tensor as $\mathbf{G} = \mathbf{I} + \mathbf{g}$ with $\|\mathbf{g}\| \ll 1$. In the absence of the growth strain tensor $\mathbf{g}$, the growth tensor reduces to a unit tensor. We assume the ribbon is made of an incompressible isotropic hyperelastic material associated with the strain energy density $\Psi = \Psi(\mathbf{A})$. The strain energy of the 3D-ribbon is

$$\mathcal{E} = \int_{\mathcal{B}_0} \Psi(\mathbf{A}) \det(\mathbf{G}) \, \mathrm{d}X_1 \mathrm{d}X_2 \mathrm{d}X_3. \tag{7}$$

The hyperelastic response is modeled with the Mooney-Rivlin constitutive law for incompressible isotropic materials. The strain energy density is given by

$$\Psi = c_1(I_1 - 3) + c_2(I_2 - 3) - p(\det(\mathbf{A}) - 1), \tag{8}$$

where $c_1$, $c_2$ are the material parameters. And $p$ is a Lagrange multiplier to be determined to enforce incompressibility. The invariants $I_1$ and $I_2$ of the right Cauchy-Green tensor $\mathbf{C} = \mathbf{A}^T \cdot \mathbf{A}$ are defined as follows

$$\begin{aligned} I_1 &= \mathrm{Tr}(\mathbf{C}) \\ I_2 &= \frac{1}{2}\left([\mathrm{Tr}(\mathbf{C})]^2 - \mathrm{Tr}(\mathbf{C}^2)\right) \end{aligned}. \tag{9}$$

Then, the second Piola-Kirchhoff stress tensor $\mathbf{S}$ for the 3D incompressible hyperelastic continuum can be derived as

$$\mathbf{S} = 2\frac{\partial \Psi}{\partial \mathbf{C}} - p\mathbf{C}^{-1} = 2(c_1 + c_2 I_1)\mathbf{I} - 2c_2\mathbf{C} - p\mathbf{C}^{-1}. \tag{10}$$

## 2.3 Asymptotic derivation of morphoelastic shell

Here, we describe the ribbon deformation based on its centerline representation, and introduce a small parameter—the thickness-to-width ratio—to reduce the 3D nonlinear elastic model to a 2D plate model similar to the von Kármán formulation.

As shown in Fig. 2(b), the mapping $\chi(\mathbf{X})$ of an arbitrary point on the deformed configuration of the ribbon can be parameterized by the centerline as

$$\chi(\mathbf{X}) = \mathbf{r}(X_3) + U_3(\mathbf{X})\mathbf{d}_3(X_3) + [X_2 + U_2(\mathbf{X})]\mathbf{d}_2(X_3) + [X_1 + U_1(\mathbf{X})]\mathbf{d}_1(X_3), \tag{11}$$

where $U_i(\mathbf{X})$ is the function to be determined that describes the local deformation of the section on the director basis. The deflection $U_1$ can be decomposed as two parts: $U_1(\mathbf{X}) = u_1(X_2, X_3) + \zeta(X_1, X_2, X_3)$, where $u_1(X_2, X_3)$ represents the deformation of the middle surface $\Omega_0$, independent of the thicknesses, and $\zeta$ accounts for the thickness-dependent contributions. By differentiating the deformation $\chi$ with respect to $\mathbf{X}$, we have the deformation gradient

$$\mathbf{F} = F_{ij}\mathbf{d}_i \otimes \mathbf{E}_j = \begin{bmatrix} 1+U_{1,X_1} & U_{1,X_2} & U_{1,X_3} + U_3\kappa_2 - (X_2+U_2)\kappa_3 \\ U_{2,X_1} & 1+U_{2,X_2} & U_{2,X_3} - U_3\kappa_1 + (X_1+U_1)\kappa_3 \\ U_{3,X_1} & U_{3,X_2} & 1+\varepsilon + U_{3,X_3} + (X_2+U_2)\kappa_1 - (X_1+U_1)\kappa_2 \end{bmatrix}. \tag{12}$$

where $U_{i,X_j} = \frac{\partial U_i}{\partial X_j}$. We can now compute the fully nonlinear strain using $\mathbf{E} = \frac{1}{2}(\mathbf{C}-\mathbf{I})$. However, to enable analytical progress in the subsequent derivation of a 1D morphoelastic ribbon model, we first introduce the following scaling analysis to simplify the system asymptotically:

(1) For the ribbon with $h \ll a \ll l_0$, we assume that the deflection $u_1$ is of the order $\eta a$, where $\eta = h/a \ll 1$ is the aspect ratio between thickness and width. The magnitudes of in-plane displacements satisfy $U_2 \sim U_3 \sim \eta^2 a$. The ribbon is nearly incompressible with Poisson's ratio $\nu = 0.5$, which implies $\zeta \sim \eta^3 a$.

(2) Following Audoly and Neukirch (2021), the scales of the axial strain, bending, and twisting curvatures of the centerline are given as $\varepsilon \sim \eta^2$, $\kappa_1 \sim \eta^2/a$, $\kappa_2 \sim \kappa_3 \sim \eta/a$.

(3) In the compatible growth, the virtual configuration coincides with the current one, i.e., $\mathbf{g}$ should be identical to the gradient of some displacement field $U_i$ with zero deflection $u_1 = 0$, then one has $g_{\alpha\beta} \sim \eta^2$, $g_{\alpha 1} \sim \eta$, $g_{1\beta} \sim \eta^3$ and $g_{11} \sim \eta^2$.

In particular, condition (1) is primarily used in the derivation of thin-shell models. This condition assumes that the in-plane elastic strains remain relatively small while the out-of-plane deflections can be large. It is consistent with the results of the Föppl–von Kármán (FvK) shell theory (Dervaux et al., 2009; Iakunin and Bonilla, 2018; Wang et al., 2024; Xu et al., 2020), in which only the nonlinear terms associated with the deflection gradients are retained in the strain expressions. Audoly and Neukirch (2021) derived condition (2) by identifying the natural orders of magnitude of the strain measures $\varepsilon$ and $\kappa_i$ of the ribbon, under the assumption that all strain measures contributions to the elastic energy are of the same order of magnitude. This approach also leads to the displacement assumptions used in condition (1). It is emphasized that these two approaches are fundamentally equivalent, both leading to a second-order approximation in terms of $\eta^2$ and resulting in a weakly nonlinear strain formulation. Condition (3) is proposed after introducing the displacement assumption to describe the case of compatible growth. This assumption is widely applied in the modeling of growing plates and shells (Iakunin and Bonilla, 2018; Wang et al., 2024; Xu et al., 2020). The detailed derivation and scaling analysis are provided in the Appendix.

With these scaling, we obtain the elastic deformation tensor

$$\mathbf{A} = \mathbf{F} \cdot \mathbf{G}^{-1} = \left( F_{ij} \mathbf{d}_i \otimes \mathbf{E}_j \right) \cdot \left( \mathbf{I} - \mathbf{g} \right) + \mathcal{O}\left( \eta^3 \right). \tag{13}$$

By retaining the dominant contributions to order $\mathcal{O}(\eta^2)$, we have the components of the asymptotic strain $\mathbf{E}$ as

$$E_{11} = \frac{1}{2}\left(U_{3,X_1}^2 + U_{2,X_1}^2 + g_{31}^2 + g_{21}^2\right) + \zeta_{,X_1} + g_{31}\left(\kappa_3 X_2 - u_{1,X_3} - U_{3,X_1}\right) - g_{21}\left(u_{1,X_2} + U_{2,X_1}\right) - g_{11}$$

$$E_{12} = E_{21} = \frac{1}{2}\left(U_{2,X_1} + u_{1,X_2} - g_{21}\right)$$

$$E_{13} = E_{31} = \frac{1}{2}\left(U_{3,X_1} + u_{1,X_3} - \kappa_3 X_2 - g_{31}\right) \quad (14)$$

$$E_{22} = U_{2,X_2} + \frac{1}{2}u_{1,X_2}^2 - g_{22}$$

$$E_{23} = E_{32} = \frac{1}{2}\left(U_{2,X_3} + U_{3,X_2} + u_{1,X_2}u_{1,X_3} + \kappa_3\left(X_1 + u_1 - X_2 u_{1,X_2}\right) - g_{23} - g_{32}\right)$$

$$E_{33} = \varepsilon + \kappa_1 X_2 - \kappa_2\left(X_1 + u_1\right) + \frac{1}{2}\left(u_{1,X_3} - \kappa_3 X_2\right)^2 + U_{3,X_3} - g_{33}$$

According to the Kirchhoff plate hypothesis that, the surface normal to the plane of the plate remains perpendicular to the plate after deformation, we have the components $E_{31} = E_{21} = 0$. The displacements $U_\alpha$ turn out to be

$$U_2(X_1, X_2, X_3) = u_2(X_2, X_3) - X_1\left(u_{1,X_2} - g_{21}\right)$$
$$U_3(X_1, X_2, X_3) = u_3(X_2, X_3) - X_1\left(u_{1,X_3} - \kappa_3 X_2 - g_{31}\right) \quad (15)$$

where $u_2(X_2, X_3)$ and $u_3(X_2, X_3)$ are the in-plane displacements of the middle surface $\Omega_0$, respectively.

Since the centerline is defined in the deformed configuration, additional constraints on the displacements need to be imposed. Submitting Eqs. (11) and (15) into (3), we have the constraints for the displacement components

$$\forall X_3 \in [0, l_0]: \int_{-a/2}^{a/2} u_1(X_2, X_3)\, dX_2 = 0,\ \int_{-a/2}^{a/2} u_2(X_2, X_3)\, dX_2 = 0,\ \int_{-a/2}^{a/2} u_3(X_2, X_3)\, dX_2 = 0. \quad (16)$$

An additional constraint regarding $u_1$ is applied by Eq. (4), i.e.,

$$\forall X_3 \in [0, l_0]: \int_{-a/2}^{a/2} X_2 u_1(X_2, X_3)\, dX_2 = 0. \quad (17)$$

This equation implies the rotation variable is the average rotation of the deformed cross section about $\mathbf{d}_3$.

Submitting Eq (15) into (14), we have the elastic strains

$$\mathbf{E} = \begin{bmatrix} e_{11} & 0 & 0 \\ 0 & e_{22} & e_{23} \\ 0 & e_{32} & e_{33} \end{bmatrix} + \mathcal{O}(\eta^3), \quad (18)$$

where $e_{\alpha\beta}$ is the 2D strain for the shell.

$$e_{22} = u_{2,X_2} + \frac{1}{2}u_{1,X_2}^2 - g_{22} - X_1\left(u_{1,X_2X_2} - g_{21,X_2}\right)$$

$$e_{32} = e_{23} = \frac{1}{2}\left(u_{3,X_2} + u_{2,X_3} + u_{1,X_2}u_{1,X_3} + \kappa_3\left(u_1 - X_2 u_{1,X_2}\right) - g_{32} - g_{23}\right) - X_1\left(u_{1,X_2X_3} - \kappa_3 - \frac{1}{2}g_{31,X_2} - \frac{1}{2}g_{21,X_3}\right). \quad (19)$$

$$e_{33} = \varepsilon + \kappa_1 X_2 - \kappa_2 u_1 + u_{3,X_3} + \frac{1}{2}\left(u_{1,X_3} - \kappa_3 X_2\right)^2 - g_{33} - X_1\left(\kappa_2 + u_{1,X_3X_3} - \kappa_{3,X_3}X_2 - g_{31,X_3}\right)$$

Here, $e_{33}$ denotes the axial strain along the length direction, $e_{22}$ represents the transverse strain in the $X_2$ direction, and $e_{23} = e_{32}$ correspond to the shear strains.

Based on the incompressibility condition $J = 1$, and $e_{\alpha\beta} \sim \mathcal{O}(\eta^2)$, $e_{\alpha\beta}^2 \sim \mathcal{O}(\eta^4)$, we have

$$\det(\mathbf{C}) - 1 = 2\text{Tr}(\mathbf{E}) + 4\left(e_{33}e_{11} + e_{22}e_{11} + e_{33}e_{22} - e_{32}^2\right) + \mathcal{O}(\eta^6) = 0. \quad (20)$$

Thus, we obtain

$$e_{11} = -(e_{33} + e_{22}) + \mathcal{O}(\eta^4). \quad (21)$$

By enforcing the plane stress condition $S_{11} = 0$, the Lagrange multiplier can be obtained as

$$p = 2c_1 + 4c_2 + 4(c_1 + c_2)e_{11} + \mathcal{O}(\eta^3). \quad (22)$$

Substituting Eqs. (21) and (22) into Eq.(10), we finally obtain the second P-K stress can be expressed as

$$\sigma_{\alpha\beta} = \frac{2}{3}Y_e\left(e_{\alpha\beta} + \delta_{\alpha\beta}e_{kk}\right) + \mathcal{O}(\eta^3), \quad (23)$$

where $Y_e$ is the equivalent Young's modulus, satisfying $Y_e = 6(c_1 + c_2)$. The elastic strain energy density can thus be expressed up to $\mathcal{O}(\eta^4)$, i.e.,

$$\begin{aligned}\Psi &= \frac{1}{2}\sigma_{\alpha\beta}e_{\alpha\beta} + \mathcal{O}(\eta^5) \\ &= \frac{2}{3}Y_e\left(e_{33}^2 + e_{22}^2 + e_{32}^2 + e_{33}e_{22}\right) + \mathcal{O}(\eta^5).\end{aligned} \quad (24)$$

In summary, the above asymptotic analysis reduces the three-dimensional nonlinear body to a 2D shell model, whose strain energy is expressed in terms of three unknown mid-surface displacement functions $u_i(X_2, X_3)$, along with the axial strain $\varepsilon$ and curvatures $\kappa_i(X_3)$ of the centerline.

## 2.4 One-dimensional ribbon model with growth

In this Section, we first reduce the above 2D model to a 1D model by introducing a small parameter (the width-to-length ratio $\gamma$) and performing a relaxation of microscopic displacements. Through a variational procedure, we derive the equilibrium equations and boundary conditions for the local displacements. Finally, we obtain a one-dimensional rod-like energy expression along with the corresponding equilibrium equations.

### 2.4.1 Dimension reduction via relaxation of microscopic displacements

To directly relate the growth strain $\mathbf{g}$ to the macroscopic strain $(\varepsilon, \kappa_1, \kappa_2, \kappa_3)$ of the ribbon, we further reduce the shell model to a 1D formulation by relaxing the microscopic displacements. This relaxation approach was first proposed by Lestringant and Audoly (2020) and has been successfully applied to the derivation of nonlinear rods model (Audoly and Lestringant, 2021), 1D ribbons model (Audoly and Neukirch, 2021; Kumar and Audoly, 2025), and higher-order tape springs model (Audoly and Lestringant, 2021).

The energy functional is written as a functional of the microscopic displacements $u_i(X_2, X_3)$, macroscopic strains $(\varepsilon, \kappa_1, \kappa_2, \kappa_3)$, and growth strain $\mathbf{g}$

$$\mathcal{E} = \int_0^{l_0} \int_{-a/2}^{a/2} \int_{-h/2}^{h/2} \Psi(u_1, u_2, u_3; \varepsilon, \kappa_1, \kappa_2, \kappa_3; \mathbf{g}) \, \mathrm{d}X_1 \mathrm{d}X_2 \mathrm{d}X_3. \tag{25}$$

Substituting the expressions of the strain components Eq. (19) into (25), we observe that the resulting terms involve derivatives with respect to $X_2$, $X_3$, and mixed derivatives $\partial(\cdot)/\partial X_2 \partial X_3$. However, for developing a 1D model, we seek a strain energy expression that depends only on the macroscopic curvature $(\varepsilon, \kappa_1, \kappa_2, \kappa_3)$ and growth strain $\mathbf{g}$. Therefore, the next step is to determine the optimal functions $u_i$ by minimizing the strain energy with a given macroscopic deformation. To this end, we introduce the following assumption: the microscopic displacements, macroscopic strains, and growth strains vary slowly along the longitudinal direction. This assumption makes the scaling assumptions that

$$\frac{\mathrm{d}^k \kappa_i}{\mathrm{d} X_3^k} = \mathcal{O}(\gamma^k), \quad \frac{\mathrm{d}^k u_i}{\mathrm{d} X_3^k} = \mathcal{O}(\gamma^k), \text{ and } \frac{\partial^k g_{ij}}{\partial X_3^k} = \mathcal{O}(\gamma^k). \tag{26}$$

where $\gamma = \dfrac{a}{l_0} \ll 1$. From Eq. (26), we can identify the relative contributions of the microscopic displacements, curvature, and growth strain to the strain energy through their presence in the strain components. For example, the component $e_{33}$ can be decomposed into three parts: a leading-order term $e_{33}^{(0)} = u_{2,X_2} + \frac{1}{2} u_{1,X_2}^2 - g_{22} - X_1 (u_{1,X_2 X_2} - g_{21,X_2}) \sim \mathcal{O}(\gamma^0)$, which does not involve derivatives with respect to $X_3$; a first-order term $e_{33}^{(1)} = \frac{1}{2}(u_{2,X_3} + u_{1,X_2} u_{1,X_3} + X_1 g_{21,X_3} - 2 X_1 u_{1,X_2 X_3}) \sim \mathcal{O}(\gamma^1)$; and a second-order term $e_{33}^{(2)} = \frac{1}{2} u_{1,X_3}^2 - X_1 u_{1,X_3 X_3} \sim \mathcal{O}(\gamma^2)$. Similarly, the strains in Eq. (19) can be expressed as a series expansion in powers of $\gamma$.

$$\begin{aligned}
e_{22} &= \underbrace{u_{2,X_2} + \frac{1}{2} u_{1,X_2}^2 - g_{22} - X_1 (u_{1,X_2 X_2} - g_{21,X_2})}_{e_{22}^{(0)} \sim \mathcal{O}(\gamma^0)} \\
e_{32} &= e_{23} = \underbrace{\frac{1}{2}(u_{3,X_2} + \kappa_3 u_1 - \kappa_3 X_2 u_{1,X_2} - g_{32} - g_{23}) + X_1 \left( \kappa_3 + \frac{1}{2} g_{31,X_2} \right)}_{e_{32}^{(0)} \sim e_{23}^{(0)} \sim \mathcal{O}(\gamma^0)} \\
&\quad + \underbrace{\frac{1}{2}(u_{2,X_3} + u_{1,X_2} u_{1,X_3} + X_1 g_{21,X_3} - 2 X_1 u_{1,X_2 X_3})}_{e_{32}^{(1)} \sim e_{23}^{(1)} \sim \mathcal{O}(\gamma^1)} \\
e_{33} &= \underbrace{\varepsilon + \kappa_1 X_2 - \kappa_2 u_1 + \frac{1}{2} \kappa_3^2 X_2^2 - \kappa_2 X_1 - g_{33}}_{e_{33}^{(0)} \sim \mathcal{O}(\gamma^0)} \\
&\quad + \underbrace{u_{3,X_3} - \kappa_3 X_2 u_{1,X_3} + \kappa_{3,X_3} X_2 X_1 + X_3 g_{31,X_3}}_{e_{33}^{(1)} \sim \mathcal{O}(\gamma^1)} + \underbrace{\frac{1}{2} u_{1,X_3}^2 - X_1 u_{1,X_3 X_3}}_{e_{33}^{(2)} \sim \mathcal{O}(\gamma^2)}
\end{aligned} \tag{27}$$

Submitting Eq. (27) into Eq.(24), one can express the strain energy as a series expansion in powers of $\gamma$.

$$\mathcal{E} = \int_0^{l_0} \int_{-a/2}^{a/2} \int_{-h/2}^{h/2} \left( \Psi^{(0)} + \Psi^{(2)} + \Psi^{(4)} \right) \mathrm{d} X_1 \mathrm{d} X_2 \mathrm{d} X_3, \tag{28}$$

where $\Psi^{(k)} = \frac{1}{2}\sigma_{\alpha\beta}e_{\alpha\beta}^{(k)}$ denotes the strain energy density, corresponding to terms of order $\gamma^k$.

Now, the objective is to determine the optimal functions $u_i$ by minimizing the strain energy with the kinematic constraints in Eqs (16) and (17) for different order of $\gamma$. Four scalar Lagrange multipliers $(f_1, f_2, f_3, f_4)$ are introduced to deal with kinematic constraints of the displacement in Eqs. (16) and (17). By Eq (25), we have the augmented total potential energy expressed as

$$\mathcal{L} = \int_0^{l_0}\int_{-a/2}^{a/2}\int_{-h/2}^{h/2}\left(\Psi(u_1,u_2,u_3;\varepsilon,\kappa_1,\kappa_2,\kappa_3;\mathbf{g}) + f_1 u_1 + f_2 u_2 + f_3 u_3 + f_4 X_2 u_1\right)\mathrm{d}X_1\mathrm{d}X_2\mathrm{d}X_3. \quad (29)$$

The equilibrium equations and boundary conditions can be derived by expanding the displacement fields $(u_1, u_2, u_3)$ in a series and substituting them into Eq. (29), followed by taking its variation and applying integration by parts.

For the sake of simplicity, we neglect the influence of higher-order terms in the longitudinal direction—specifically, the second and fourth-order terms in the strain energy—and retain only the leading-order term in $\mathcal{O}(\gamma^0)$. By applying the variational method, we obtain the Euler–Lagrange equations

$$\frac{Y_e h}{3}\left(g_{23} + g_{32} - \kappa_3 u_1 - u_{3,X_2} + \kappa_3 X_2 u_{1,X_2}\right)_{,X_2} + f_3 = 0, \quad (30)$$

$$\frac{Y_e h}{3}\left(2g_{33} + 4g_{22} - 2\varepsilon - 2X_2\kappa_1 + 2\kappa_2 u_1 - X_2^2\kappa_3^2 - 4u_{2,X_2} - 2u_{1,X_2}^2\right)_{,X_2} + f_2 = 0, \quad (31)$$

$$\frac{Y_e h^3}{18}\left[\kappa_2 + 2u_{1,X_2 X_2} - 2g_{21,X_2}\right]_{,X_2 X_2} + Y_e h \kappa_2^2 u_1 + Y_e h \kappa_2\left(g_{33} - \varepsilon - \kappa_1 X_2 - \frac{\kappa_3^2 X_2^2}{2} + \frac{f_1 + X_2 f_4}{Y_e h \kappa_2}\right) = 0. \quad (32)$$

The boundary conditions are given by

$$g_{23} + g_{32} - \kappa_3 u_1 - u_{3,X_2} + \kappa_3 X_2 u_{1,X_2} = 0 \text{ for } \forall X_3 \in [0, l_0], X_2 = \pm/2, \quad (33)$$

$$2g_{33} + 4g_{22} - 2\varepsilon - 2X_2\kappa_1 + 2\kappa_2 u_1 - X_2^2\kappa_3^2 - 4u_{2,X_2} - 2u_{1,X_2}^2 = 0 \text{ for } \forall X_3 \in [0, l_0], X_2 = \pm a/2, \quad (34)$$

$$\kappa_2 + 2u_{1,X_2 X_2} - 2g_{21,X_2} = 0 \text{ for } \forall X_3 \in [0, l_0], X_2 = \pm a/2, \quad (35)$$

$$\left(\kappa_2 + 2u_{1,X_2 X_2} - 2g_{21,X_2}\right)_{,X_2} = 0 \text{ for } \forall X_3 \in [0, l_0], X_2 = \pm a/2. \quad (36)$$

When the growth strain $g_{ij}$ is neglected, the model reduces to the nonlinear ribbon model previously derived by Audoly and Lestringant (2021). This nonlinear model accurately predicts the torsional behavior of ribbons and has been validated by finite element simulation and experimental results in previous work (Liu et al., 2025).

Integrating Eqs. (30) and (31) with $X_2$ and using the boundary conditions Eqs. (33), (34) and the kinematic constraints of the displacement Eqs. (16), we have $f_3 = f_2 = 0$. An analogous procedure is carried out for Eq. (84), giving that

$$f_1 = \frac{Y_e h \kappa_2}{24}\left(24\varepsilon + a^2 \kappa_3^2\right) - \frac{Y_e h \kappa_2}{a} \int_{-a/2}^{a/2} g_{33}(X_2, X_3) \mathrm{d}X_2, \tag{37}$$

and

$$f_4 = Y_e h \kappa_1 \kappa_2 - Y_e h \kappa_2 \frac{12}{a^3} \int_{-a/2}^{a/2} X_2 g_{33}(X_2, X_3) \mathrm{d}X_2. \tag{38}$$

Equation (32) governs the equilibrium of $u_1(X_2, X_3)$ on the mid-surface $\Omega_0$. It indicates that the displacement magnitude is influenced by the growth strain components $g_{21}$, $g_{33}$. Once the corresponding growth components are prescribed, the general solution for the displacement $u_1$ can be obtained. The coefficients in the solution can be determined by the boundary conditions Eqs. (35), (36) and the displacement constraints (Eq. (16) and (17)). Substituting the solution of Eq. (32) into Eqs. (30) and (31), we then obtain the solutions for $u_3$ and $u_2$, respectively.

Through the relaxation approach, we eliminate the dependence of the displacement on the transverse coordinate $X_2$. Notably, the resulting displacement field is derived via a variational method and is thus asymptotic exact for the model. Substituting the obtained displacement field into Eq. (27) yields the corresponding strain expressions. According to Eq. (24), and by integrating it over the thickness and width, we derive an energy expression that depends solely on the macroscopic strains $(\varepsilon, \kappa_1, \kappa_2, \kappa_3)$ and growth strains $\mathbf{g}$, that is

$$\mathcal{E} = \int_0^{l_0} W(\varepsilon, \kappa_1, \kappa_2, \kappa_3; \mathbf{g}) \, \mathrm{d}X_3. \tag{39}$$

This equation means that once the growth tensor $\mathbf{g}$ is prescribed, the strain energy can ultimately be expressed in terms of the macroscopic strains $(\varepsilon, \kappa_1, \kappa_2, \kappa_3)$ along the centerline. This formulation resembles the classical Kirchhoff rod model. However, it differs in that the present model incorporates weak geometric nonlinearities and is asymptotically accurate up to second order in the small parameter $\eta$. Furthermore, by assuming slow variations of curvature and displacements along the longitudinal direction, the model can capture weakly nonlinear behavior and finite rotations induced by growth.

### 2.4.2 Balance and constitutive laws of the rod

In the Kirchhoff rod theory, the stress acting on the cross-section at $X_3$ gives rise to a resultant force $\mathbf{N}(X_3) = N_i \mathbf{d}_i(X_3)$ and resultant moment $\mathbf{M}(X_3) = M_i \mathbf{d}_i(X_3)$ attached to the centerline $\mathbf{r}(X_3)$. Neglecting the inertial effects leads to the 1D rod Kirchhoff equations

$$\begin{aligned} \mathbf{N}' + \mathbf{p} &= \mathbf{0} \\ \mathbf{M}' + \mathbf{r}' \times \mathbf{N} + \mathbf{q} &= \mathbf{0} \end{aligned} \qquad (40)$$

where $\mathbf{p}$ and $\mathbf{q}$ are the external forces (i.e., the gravity of the ribbon) and torques per unit length. The internal moments and force are given by

$$\begin{aligned} M_i &= \frac{\partial W(\varepsilon, \kappa_1, \kappa_2, \kappa_3; \mathbf{g})}{\partial \kappa_i} \\ N_3 &= \frac{\partial W(\varepsilon, \kappa_1, \kappa_2, \kappa_3; \mathbf{g})}{\partial \varepsilon} \end{aligned} \qquad (41)$$

Since we assume that the ribbon is non-shareable, the forces $N_1 = N_2 = 0$.

### 2.4.3 Non-dimensionalization

For simplicity, we introduce the following non-dimensional variables

$$K_i = \kappa_i a, \quad \Pi = \frac{W}{Y_e h a}, \quad m_i = \frac{M_i}{Y_e h a^2}, \quad n_i = \frac{N_i}{Y_e h a}. \qquad (42)$$

where $\Pi$ represents the non-dimensional strain energy, $K_i$ is the non-dimensional curvatures. Correspondingly, the non-dimensional force $n_i$, and moment $m_i$ can be

given by

$$n_3 = \frac{\partial \Pi}{\partial \varepsilon} \quad \text{and} \quad m_i = \frac{\partial \Pi}{\partial K_i}. \tag{43}$$

## 3 Morphological patterns of leaves and petals

Here, we use the 1D morphoelastic ribbon model and specified growth tensors to investigate the 3D morphologies exhibited by slender plant leaves and flower petals, including saddle-bending, twisting, and helical configurations.

### 3.1 Saddle-bending configuration

Repeated opening and closure of the flower are driven by asymmetric cell expansion between the adaxial and abaxial sides of the mesophyll in response to environmental cues such as temperature, light, or humidity (van Doorn and Kamdee, 2014). This implies that external stimuli induce curvature through differential growth. We take the incremental growth to be purely $X_3$ direction with a gradient along the $X_1$ axis,

$$\mathbf{G} = \begin{bmatrix} 1 & 0 & 0 \\ 0 & 1 & 0 \\ 0 & 0 & 1 + \xi X_1/a \end{bmatrix}, \tag{44}$$

where $\xi$ denotes the growth strain. The growth strain $g_{33}$ varies linearly along the $X_1 \in [-h/2, h/2]$ axis, reaching its maximum at the surfaces. It is important to note that at $X_1 = 0$, i.e., on the mid-surface $\Omega_0$, $g_{33} = 0$, indicating that the growth strain does not directly affect the mid-surface. As a result, in ribbon-shaped leaves, the differential growth across the thickness leads to a bending configuration with $\kappa_2 \neq 0$, and $\kappa_3 = 0$; as shown in Fig. 3.

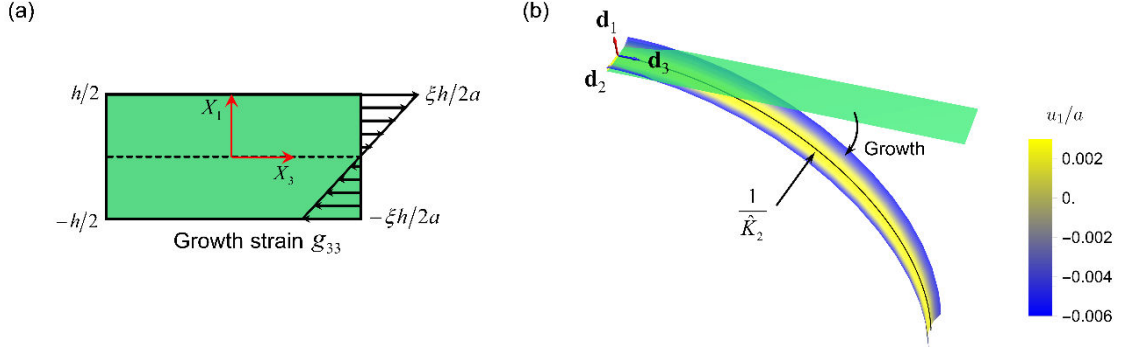

*Fig. 3. A linear growth gradient in the $X_1$ direction generates a ribbon with intrinsic curvature $\hat{K}_2$. (a) Growth gradient in the thickness direction, (b) Saddle-bending configuration.*

Substituting $\kappa_3 = 0$ and Eq. (44) into (32) yields a fourth-order ordinary differential equation governing the out-of-plane deflection

$$u_{1,X_2X_2X_2X_2} + \beta^2 u_1 = 0 . \tag{45}$$

where $\beta^2 = \dfrac{9\kappa_2^2}{h^2}$. The general solution of Eq. (45) can be expressed as a linear combination of $\cos(qX_2)\cosh(qX_2)$, $\cosh(qX_2)\sin(qX_2)$, $\cos(qX_2)\sinh(qX_2)$, and $\sin(qX_2)\sinh(qX_2)$, i.e.,

$$\tilde{u}_1 = C_1 \cos(qX_2)\cosh(qX_2) + C_2 \cosh(qX_2)\sin(qX_2) \\ + C_3 \cos(qX_2)\sinh(qX_2) + C_4 \sin(qX_2)\sinh(qX_2) . \tag{46}$$

where $q = \sqrt{\beta/2} = \sqrt{\dfrac{3|\kappa_2|}{2h}}$. The coefficients can be determined by applying the boundary conditions Eqs. (35), (36), and the integral constraint specified in Eqs. (16) and (17). They are

$$\begin{aligned}
C_1 &= \kappa_2 \frac{\cosh\left(\frac{qa}{2}\right)\sin\left(\frac{qa}{2}\right) - \cos\left(\frac{qa}{2}\right)\sinh\left(\frac{qa}{2}\right)}{2q^2(\sin(qa)+\sinh(qa))\operatorname{sign}(\kappa_2)} \\
C_2 &= C_3 = 0 \\
C_4 &= -\kappa_2 \frac{\cosh\left(\frac{qa}{2}\right)\sin\left(\frac{qa}{2}\right) + \cos\left(\frac{qa}{2}\right)\sinh\left(\frac{qa}{2}\right)}{2q^2(\sin(qa)+\sinh(qa))\operatorname{sign}(\kappa_2)}
\end{aligned} \tag{47}$$

Substituting Eq. (47) into Eq. (46) yields the expression of $u_1$ for the saddle-bending ribbons,

$$u_1 = \frac{\cos(qX_2)\cosh(qX_2)\left(\cosh\left(\frac{qa}{2}\right)\sin\left(\frac{qa}{2}\right) - \cos\left(\frac{qa}{2}\right)\sinh\left(\frac{qa}{2}\right)\right)\kappa_2}{2q^2(\sin(qa) + \sinh(qa))\text{sign}(\kappa_2)}$$
$$- \frac{\sin(qX_2)\sinh(qX_2)\left(\cosh\left(\frac{qa}{2}\right)\sin\left(\frac{qa}{2}\right) + \cos\left(\frac{qa}{2}\right)\sinh\left(\frac{qa}{2}\right)\right)\kappa_2}{2q^2(\sin(qa) + \sinh(qa))\text{sign}(\kappa_2)}. \quad (48)$$

This result illustrates the distribution of the deflection $u_1$ along the $X_2$ axis under different bending curvatures $\kappa_2$. Based on the expression of $u_1$, the corresponding solutions for $u_2$ and $u_3$ can be derived. Subsequently, by applying Eqs. (19), (23) and (24), the expression for the strain energy per unit length can be determined

$$W = \frac{Y_e ah}{2}\varepsilon^2 + \frac{Y_e a^3 h}{24}\kappa_1^2 + \frac{Y_e ah^3}{18}(\kappa_2 + \xi/a)^2 + V(\kappa_2), \quad (49)$$

where

$$V(\kappa_2) = \frac{Y_e h^{7/2}\sqrt{\kappa_2(\kappa_2 + 2\xi/a)}}{18\sqrt{6}} \frac{\cos(aq) - \cosh(aq)}{(\sin(aq) + \sinh(aq))} \quad (50)$$

Notably, the term $V(q)$ arises from the variation in deflection induced by the bending of the ribbon. By Eq. (49), we conclude that the morphoelastic ribbon acquires an intrinsic curvature $\hat{K}_2$ due to a growth gradient along the $X_1$ direction, as shown in Fig. 3(b).

Fig. 4(a) illustrates the variation of the normalized strain energy given by Eq. (49) with respect to $K_2$ for different values of $\xi$, where the dimensionless energy $\Pi = W/Yha$. In this case, we set $\eta = 0.05$, and $\varepsilon = K_1 = 0$. It can be observed that when the growth strain occurs only along the $X_3$ direction and exhibits a linear gradient in the $X_1$ direction, the strain energy attains a minimum at a specific curvature $\hat{K}_2$. Furthermore, as the value of $\xi$ increases, the location of the energy minimum shifts towards a larger value of $\hat{K}_2$.

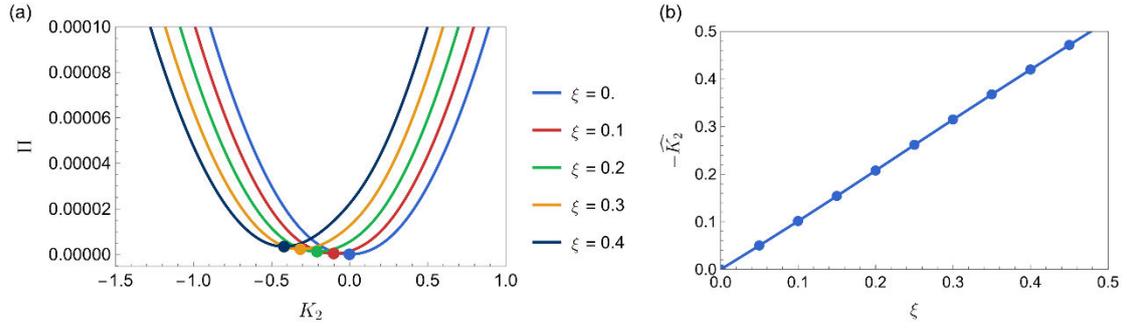

*Fig. 4. Variation of normalized strain energy with $K_2$ for different values of $\xi$. (a) Normalized strain energy for $\eta = 0.05$ and $\varepsilon = K_1 = 0$. (b) The intrinsic curvature increases monotonically with growth strain.*

As shown in Fig. 4(b), the value of intrinsic curvature $\hat{K}_2$ increases linearly with the growth strain amplitude $\xi$, under the condition $\eta = 0.05$, and $\varepsilon = K_1 = 0$. It indicates that there is a linear correlation between the growth strain and the resulting curvature. As indicated by Eq. (48), the 1D deflection $u_1/a$ across the $X_2$-axis depends on the intrinsic curvature $\hat{K}_2$ and the thickness-to-width ratio $\eta$. The morphological evolution of the ribbon during the bending process is illustrated in Fig. 5. Figures 5(a) and (b) correspond to ribbons with thickness-to-width ratios of 0.05 and 0.1, respectively. The curvature of the ribbon remains consistent across different thickness-to-width ratios as the imposed growth strain increases from 0 to 0.3. This invariance arises because, as shown in Eq. (49), the normalized bending curvature is independent of the thickness-to-width ratio. Nevertheless, for a given growth strain, notable differences emerge in the cross-sectional shapes of the ribbons. At the initial stage, the ribbon remains flat without observable deformation. With a further increase in $\xi$, the ribbon progressively transitions into a saddle-like configuration.

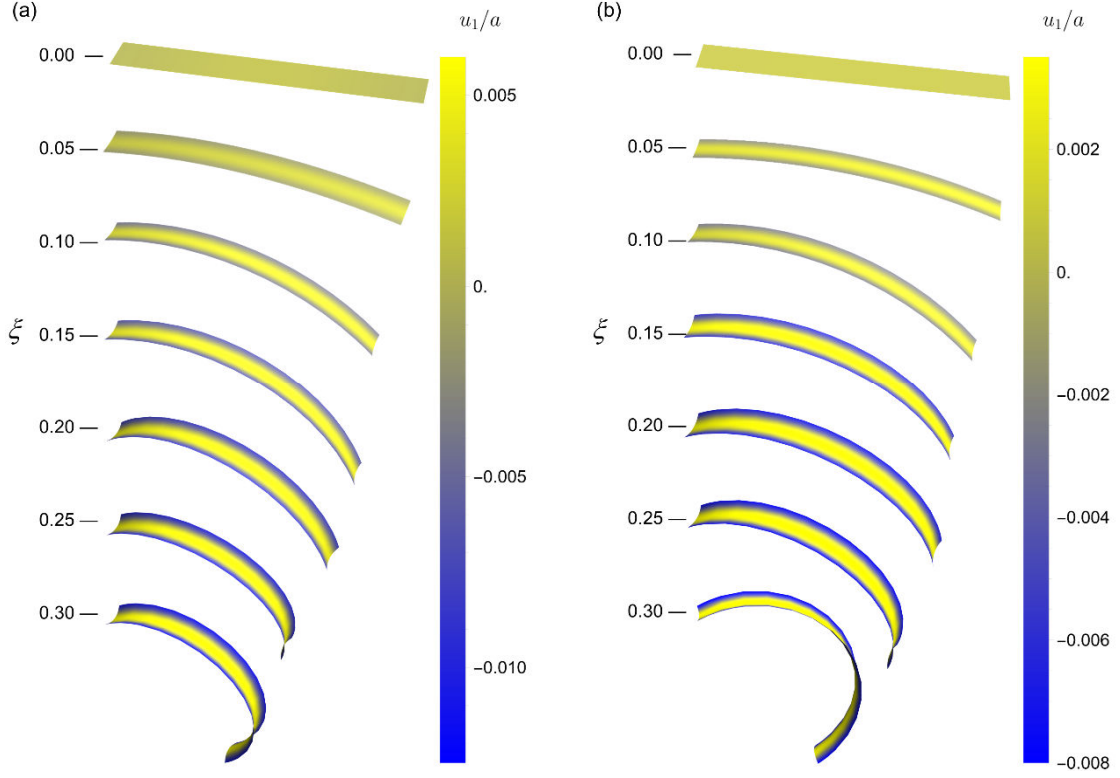

*Fig. 5 Morphological evolution of the saddle bending ribbon: (a) $\eta = 0.05$ and (b) $\eta = 0.1$.*

## 3.2 Twisting configuration

Helical growth is another typical growth pattern observed in nature. It is associated with excessive growth in the central region of the leaf, such as tip growth effects, which necessitate a twisting deformation to release the mismatch stresses induced by differential growth. Inspired by Huang et al. (2018), we consider a growth strain in the longitudinal direction that exhibits a gradient along the $X_2$ - axis

$$\mathbf{G} = \begin{bmatrix} 1 & 0 & 0 \\ 0 & 1 & 0 \\ 0 & 0 & 1+\xi(2X_2/a)^2 \end{bmatrix}. \quad (51)$$

The growth strain $g_{33}$ monotonically increases from zero at the center to a maximum value $\xi$ at the edge $X_2 = \pm a/2$. Substituting Eq. (51) into Eq. (32) yields a simplified form of the governing equation

$$u_{1,X_2X_2X_2X_2} + \beta^2 u_1 = \frac{3\kappa_2}{8a^2h^2}(12X_2^2 - a^2)(a^2\kappa_3^2 - 8\xi). \quad (52)$$

The general solution of Eq. (52) is the sum of the homogeneous solution to the homogeneous Eq. (45) and a particular solution, i.e.,

$$u_1 = \tilde{u}_1 + u_1^*. \tag{53}$$

In this case, a particular solution can be readily obtained as

$$u_1^* = \frac{(12X_2^2 - a^2)(a^2\kappa_3^2 - 8\xi)}{24a^2\kappa_2}. \tag{54}$$

The expression for $\tilde{u}_1$ has already been provided in Eq. (46). Here, we can determine the coefficients by applying the boundary conditions and the integral constraint again. They are

$$\begin{aligned}
C_1 &= \frac{\cosh\left(\frac{qa}{2}\right)\sin\left(\frac{qa}{2}\right) - \cos\left(\frac{qa}{2}\right)\sinh\left(\frac{qa}{2}\right)}{2q^2(\sin(qa) + \sinh(qa))\operatorname{sign}(\kappa_2)}\left(\frac{a^2\kappa_2^2 + 2a^2\kappa_3^2 - 16\xi}{a^2\kappa_2}\right) \\
C_2 &= C_3 = 0 \\
C_4 &= -\frac{\cosh\left(\frac{qa}{2}\right)\sin\left(\frac{qa}{2}\right) + \cos\left(\frac{qa}{2}\right)\sinh\left(\frac{qa}{2}\right)}{2q^2(\sin(qa) + \sinh(qa))\operatorname{sign}(\kappa_2)}\left(\frac{a^2\kappa_2^2 + 2a^2\kappa_3^2 - 16\xi}{a^2\kappa_2}\right)
\end{aligned}. \tag{55}$$

Substituting the coefficient expressions from Eq. (55) into Eq. (53) yields the solution corresponding to the imposed growth strain

$$\begin{aligned}
u_1 &= \frac{\cosh\left(\frac{qa}{2}\right)\sin\left(\frac{qa}{2}\right) - \cos\left(\frac{qa}{2}\right)\sinh\left(\frac{qa}{2}\right)}{2q^2(\sin(qa) + \sinh(qa))\operatorname{sign}(\kappa_2)}\left(\frac{a^2\kappa_2^2 + 2a^2\kappa_3^2 - 16\xi}{a^2\kappa_2}\right)\cos(qT)\cosh(qT) \\
&\quad - \frac{\cosh\left(\frac{qa}{2}\right)\sin\left(\frac{qa}{2}\right) + \cos\left(\frac{qa}{2}\right)\sinh\left(\frac{qa}{2}\right)}{2q^2(\sin(qa) + \sinh(qa))\operatorname{sign}(\kappa_2)}\left(\frac{a^2\kappa_2^2 + 2a^2\kappa_3^2 - 16\xi}{a^2\kappa_2}\right)\sin(qT)\sinh(qT) \\
&\quad + \frac{(12X_2^2 - a^2)(a^2\kappa_3^2 - 8\xi)}{24a^2\kappa_2}
\end{aligned} \tag{56}$$

Subsequently, by Eqs. (19), (23) and (24), the expression for the strain energy per unit length can be determined as

$$\begin{aligned}
W &= \frac{Y_e a h}{2}\left(\varepsilon + \frac{a^2\kappa_3^2}{24}\right)^2 + \frac{Y_e a^3 h}{24}\kappa_1^2 + \frac{Y_e a h^3}{18}\frac{(\kappa_2^2 + \kappa_3^2)^2}{\kappa_2^2} \\
&\quad + \frac{Y_e a h}{2}\left(\frac{(\xi - 6\varepsilon)}{9} + \frac{64h^2\xi}{9a^4\kappa_2^2} - \frac{8h^2(\kappa_2^2 + 2\kappa_3^2)}{9a^2\kappa_2^2} - \frac{a^2\kappa_3^2}{36}\right)\xi + V(\kappa_2, \kappa_3)
\end{aligned}, \tag{57}$$

where

$$V(\kappa_2, \kappa_3) = \frac{Y_e h^{7/2}\left(a^2(\kappa_2^2 + 2\kappa_3^2) - 16\xi\right)^2(\cos(qa) - \cosh(qa))}{18\sqrt{6}a^4\kappa_2^{5/2}(\sin(qa) + \sinh(qa))}. \tag{58}$$

As $K_2 \sim \eta \ll 1$, we retain up to the second-order terms of $K_2$ in the normalized strain energy Eq. (57), leading to the simplified expression

$$\Pi = \frac{1}{2}\left(\varepsilon + \frac{K_3^2}{24}\right)^2 + \frac{K_1^2}{24} + \frac{K_3^4}{1440} + \frac{K_3^2}{360}(20\eta^2 - 9\xi) - \frac{\varepsilon\xi}{3} + \frac{\xi^2}{10} + K_2^2\left(\frac{K_3^2 + 60\eta^2 - 8\xi}{1440} - \frac{(K_3^2 - 8\xi)^2}{80640\eta^2}\right). \quad (59)$$

The 3D contour of the normalized strain energy as a function of $K_2$ and $K_3$ is shown in Fig. 6(a), under the conditions of $\eta = 0.1$ and $\xi = 0.03$. As shown in Fig. 6(a), there are three energy extrema: the origin, and two nontrivial points with $K_2 = 0$ and $K_3 = \pm \hat{K}_3$. The origin corresponds to the trivial solution, which is not of interest in this context. The extremum at $K_2 = 0$ and $K_3 \neq 0$ suggests that, under the stimulus of a growth strain $g_{33} = \xi(2X_2/a)^2$, the grown leaf or petal may adopt a twisting configuration as its final shape, as shown in Fig. 6(b).

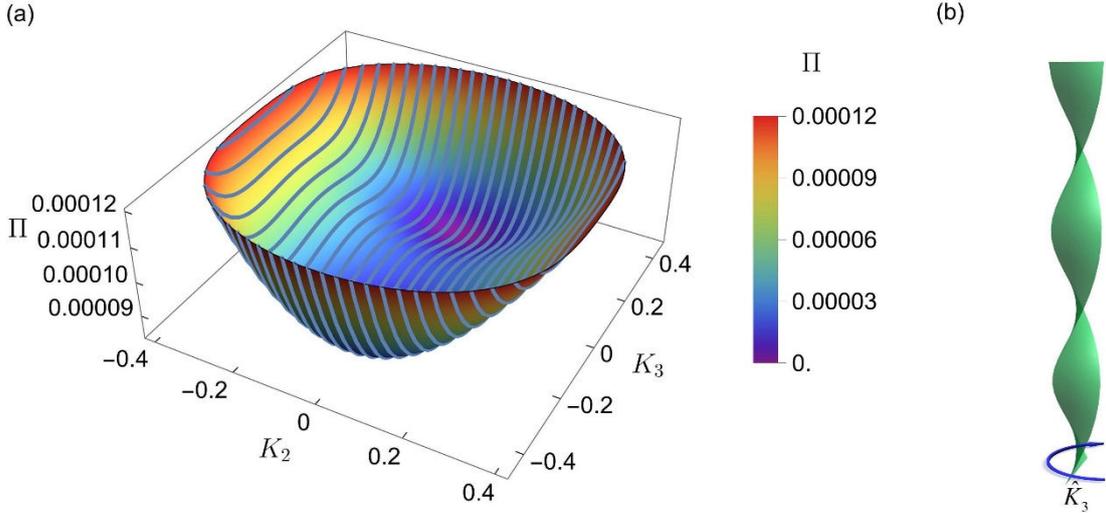

*Fig. 6 (a) Contour plot of the normalized strain energy as a function of $K_2$ and $K_3$. (b) Twisting configuration.*

The variation of the normalized strain energy $\Pi$ of the ribbons for different values of $\eta$ is shown in Fig. 7, with the parameters $K_2 = 0$ and $\xi = 0.03$. The disk markers indicate the locations of the intrinsic curvature $\hat{K}_3$. As the value of $\eta$ increases, the value of $\hat{K}_3$ gradually decreases, indicating that ribbons with larger thickness-to-width ratios exhibit lower levels of twisting under the same conditions.

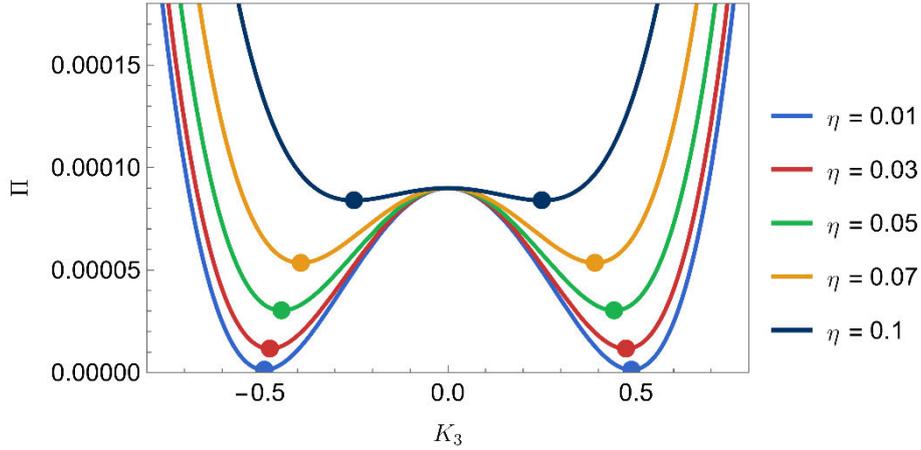

*Fig. 7 Variation of normalized strain energy with $K_3$ for different values of $\eta$.*

The variation of the normalized strain energy $\Pi$ of the ribbons for different values of $\xi$ is shown in Fig. 8, with the parameters $K_2 = 0$ and $\eta = 0.05$. The disk markers indicate the locations of $\hat{K}_3$. As the value of $\xi$ increases, the origin remains an unstable critical point, while the value of $\hat{K}_3$ progressively increases.

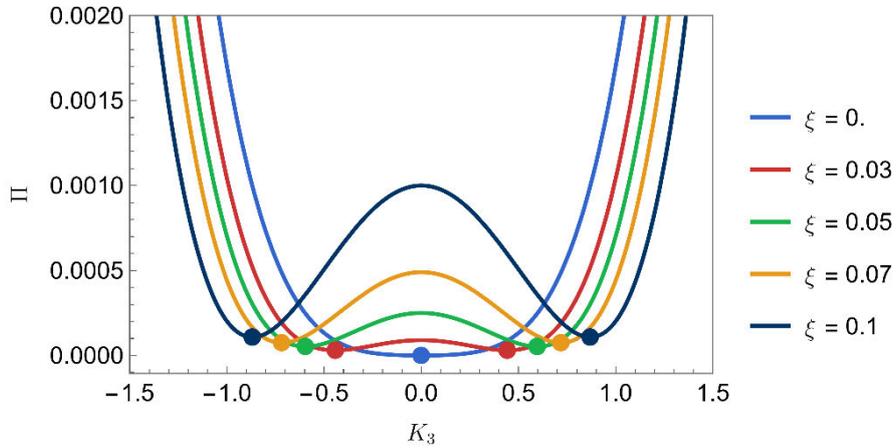

*Fig. 8 Variation of normalized strain energy with $K_3$ for different values of $\xi$.*

To determine the value of the intrinsic curvature $\hat{K}_3$, we employ the developed 1D Kirchhoff ribbon equations, i.e., Eq (41). We begin by projecting the balance equations onto the local orthonormal basis $(\mathbf{d}_1, \mathbf{d}_2, \mathbf{d}_3)$, which yields six scalar equilibrium equations

$$N_1' - N_2\kappa_3 + N_3\kappa_2 = 0$$
$$N_2' - N_3\kappa_1 + N_1\kappa_3 = 0$$
$$N_3' - N_1\kappa_2 + N_2\kappa_1 = 0$$
$$M_1' - M_2\kappa_3 + M_3\kappa_2 - (1+\varepsilon)N_2 = 0 \quad (60)$$
$$M_2' + M_1\kappa_3 - M_3\kappa_1 + (1+\varepsilon)N_1 = 0$$
$$M_3' - M_1\kappa_2 + M_2\kappa_1 = 0$$

By Eq. (43), the normalized force and moment acting on the ribbon's centerline under the influence of the growth tensor described by Eq. (51) can be derived as

$$m_1 = \frac{\partial \Pi}{\partial K_1} = \frac{K_1}{12}$$
$$m_2 = \frac{\partial \Pi}{\partial K_2} = K_2 \left( \frac{K_3^2 + 60\eta^2 - 8\xi}{720} - \frac{(K_3^2 - 8\xi)^2}{40320\eta^2} \right)$$
$$m_3 = \frac{\partial \Pi}{\partial K_2} = K_3 \left( \frac{\varepsilon}{12} + \frac{\eta^2}{9} - \frac{\xi}{20} + K_2^2 \left( \frac{1}{720} + \frac{\xi}{2520\eta^2} \right) \right) + K_3^3 \left( \frac{1}{160} - \frac{K_2^2}{20160\eta^2} \right) \quad (61)$$
$$n_3 = \frac{\partial \Pi}{\partial \varepsilon} = \frac{K_3^2}{24} + \varepsilon - \frac{\xi}{3}$$

The equilibrium configuration of the ribbon is assumed to be in a stress-free state. To determine the value of $\hat{K}_3$, we solve the above equilibrium equations along with the corresponding boundary conditions. For the twisting configuration, the macroscopic curvatures are $K_1 = K_2 = 0$, while $K_3 = \hat{K}_3$. Consequently, the projected equilibrium equations can be simplified to

$$N_3' = M_3' = 0 \quad (62)$$

with boundary conditions

$$N_3 = M_3 = 0, \text{ at } X_3 = 0 \text{ or } l_0 \quad (63)$$

Substituting the constitutive laws Eq. (61) into (62), we can obtain

$$\hat{K}_3 = \pm 2\sqrt{2}\sqrt{\xi - 5\eta^2} . \quad (64)$$

The above equation (64) implies that, as the value of $\xi$ increases, the flat ribbon gradually becomes unstable and transitions into a twisting configuration at a critical point. For the twisting ribbon with a given thickness-to-width ratio where $|K_3| \neq 0$, we can then determine that the critical buckling point occurs at $\xi^* = 5\eta^2$ by Eq. (64). Plots of the variation of $K_3$ with $\xi$ for different values of $\eta$ are given in Fig. 9. When the value of $\xi$ is smaller than the critical value $\xi^*$, $K_3 = 0$, and the ribbon remains in a flat state. When $\xi > \xi^*$, the ribbon transitions from a flat configuration to a twisting configuration. The growth threshold $\xi^*$ for the helical transition increases with the ratio of thickness to width, as indicated by the pentagram symbols in Fig. 9.

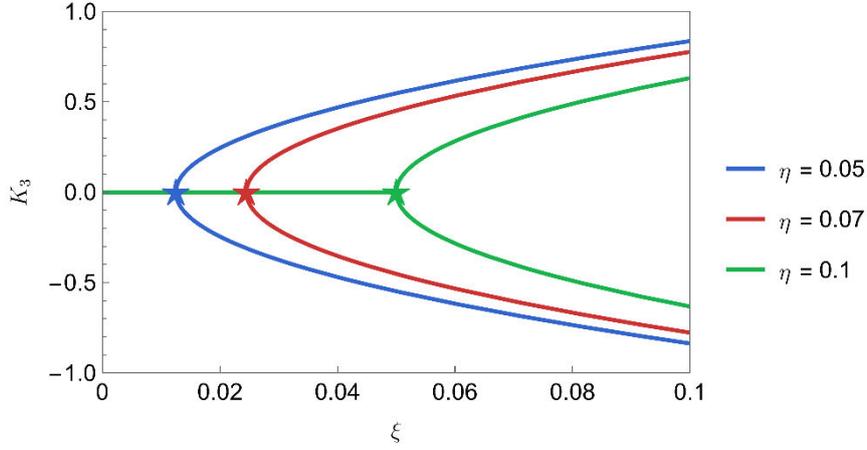

Fig. 9 Variation of $K_3$ with the growth strain $\xi$ for different values of $\eta$.

## 3.3 Helical Configuration

Here, we continue to consider the growth strain defined in Eq. (51). By numerically solving the governing equations in conjunction with appropriate boundary conditions, we interpret the emergence of helical configurations as a bifurcation phenomenon induced by differential growth.

To relate the local basis to the global basis and solve the equilibrium equations, a unit quaternion $\mathbf{q}(X_3) = (q_0(X_3), q_1(X_3), q_2(X_3), q_3(X_3))$ with $q_0^2 + q_1^2 + q_2^2 + q_3^2 = 1$ is introduced (Healey and Mehta, 2005; Yu and Hanna, 2019), which gives

$$\begin{Bmatrix} \mathbf{d}_1 \\ \mathbf{d}_2 \\ \mathbf{d}_3 \end{Bmatrix} = 2 \begin{bmatrix} q_0^2 + q_1^2 - \dfrac{1}{2} & q_1 q_2 + q_0 q_3 & q_1 q_3 - q_0 q_2 \\ q_1 q_2 - q_0 q_3 & q_0^2 + q_2^2 - \dfrac{1}{2} & q_2 q_3 + q_0 q_1 \\ q_1 q_3 + q_0 q_2 & q_2 q_3 - q_0 q_1 & q_0^2 + q_3^2 - \dfrac{1}{2} \end{bmatrix} \begin{Bmatrix} \mathbf{E}_1 \\ \mathbf{E}_2 \\ \mathbf{E}_3 \end{Bmatrix} = [\mathbf{Q}] \begin{Bmatrix} \mathbf{E}_1 \\ \mathbf{E}_2 \\ \mathbf{E}_3 \end{Bmatrix}. \qquad (65)$$

where $\mathbf{Q}$ is a rotation matrix. Substituting $\mathbf{d}_3$ from Eq. (65) into Eq.(1), the kinematic equations of the centerline can be expressed by the quaternion components as

$$r_1' = 2(1+\varepsilon)(q_1 q_3 + q_0 q_2), r_2' = 2(1+\varepsilon)(q_2 q_3 - q_0 q_1), r_3' = 2(1+\varepsilon)\left(q_0^2 + q_3^2 - \frac{1}{2}\right), \quad (66)$$

where $r_i$ $(i=1,2,3)$ is the component of the position vector $\mathbf{r}(X_3)$, and $\mathbf{r}(X_3) = r_1 \mathbf{E}_1 + r_2 \mathbf{E}_2 + r_3 \mathbf{E}_3$. By taking the derivative of Eq. (65) to $X_3$ and substituting it into Eq. (5), the kinematic equations governing the cross-sectional orientation can be reformulated

$$\begin{aligned} q_0' &= \frac{1}{2}(-q_1 \kappa_1 - q_2 \kappa_2 - q_3 \kappa_3) \\ q_1' &= \frac{1}{2}(q_0 \kappa_1 - q_3 \kappa_2 + q_2 \kappa_3) \\ q_2' &= \frac{1}{2}(q_3 \kappa_1 + q_0 \kappa_2 - q_1 \kappa_3) \\ q_3' &= \frac{1}{2}(-q_2 \kappa_1 + q_1 \kappa_2 + q_0 \kappa_3) \end{aligned} \quad (67)$$

Incorporating the constitutive relation from Eq. (61), the combined system of Eqs.(60), (66) and (67) provide 13 first-order ordinary differential equations for the ribbon, involving thirteen dependent variables $(\varepsilon, \kappa_i, r_i, q_0, q_1, q_2, q_3)$ $(i=1,2,3)$.

At $X_3 = 0$, the ribbon is assumed to be clamped, corresponding to the following seven boundary conditions:

$$\begin{aligned} r_1(X_3 = 0) &= r_2(X_3 = 0) = r_3(X_3 = 0) = 0 \\ q_0(X_3 = 0) &= 1, \, q_1(X_3 = 0) = q_2(X_3 = 0) = q_3(X_3 = 0) = 0 \end{aligned} \quad (68)$$

At $X_3 = l_0$, the leaf/petal is assumed to be in a free state, which leads to the six boundary conditions:

$$\begin{aligned} N_1(X_3 = 0) &= N_2(X_3 = 0) = N_3(X_3 = 0) = 0 \\ M_1(X_3 = 0) &= M_2(X_3 = 0) = M_3(X_3 = 0) = 0 \end{aligned} \quad (69)$$

With these boundary conditions, the governing equations form a well-posed two-point boundary value problem (BVP). This BVP is numerically solved using the continuation software package *COCO* (Continuation Core and Toolboxes) (Ahsan et al., 2022; Dankowicz and Schilder, 2011), implemented in MATLAB (R2022a, MathWorks). The system is discretized using the orthogonal collocation method, which transforms the differential equations into a set of algebraic equations. The package *COCO* employs a combination of continuation algorithms and Newton iteration techniques to trace solution branches efficiently. Here, the growth strain $\xi$ is selected as the sole continuation parameter. Notably, *COCO* is capable of detecting bifurcation points, enabling the tracking of solution paths across multiple branches and identifying the onset of instabilities.

The bifurcation diagram for the ribbon with $\eta = 0.1$ is shown in Fig. 10. The evolutions of twisting and bending curvatures as functions of the growth strain are shown in Fig. 10(a) and (b), respectively. It is noted that the curvature shown here is defined as the average curvature $\bar{K}_i$, since, in the numerical results, the curvature along the longitudinal direction $X_1$ may vary spatially

$$\bar{K}_i = \frac{1}{n}\sum_{j=1}^{n} K_i^j. \tag{70}$$

where $n$ and $K_i^j$ denote the number of discrete points along the longitudinal direction and the corresponding curvature values, respectively. The red solid lines in Fig. 10(a) represent theoretical predictions by Eq. (64), while the square and circular markers denote numerical results obtained using the continuation software *COCO*. As the growth parameter $\xi$ increases, the first bifurcation occurs at $\xi \approx 0.05$, marking the transition from a flat configuration to a twisted one. The sign of $\bar{K}_3$ indicates the handedness of the twist, representing opposite chirality. The numerical results show good agreement with the theoretical predictions. As $\xi$ further increases to $\xi \approx 0.53$, a secondary bifurcation is observed in the numerical results, indicating a transition from a twisting configuration to a helical one. In this stage, both $\bar{K}_2$ and $\bar{K}_3$ increase progressively with $\xi$, corresponding to a continuously increasing degree of helicity.

Fig. 10(c) and (d) show the curvature distribution along the longitudinal direction for the twisting and helical configurations, respectively. In the twisting configuration, the ribbon exhibits intrinsic twist $\hat{K}_3$ while the principal bending curvatures remain essentially zero ($\hat{K}_2 = \hat{K}_1 = 0$), as shown in Fig. 10(c). In the helical configuration, the ribbon possesses both intrinsic curvature and torsion. Due to the small thickness-to-width ratio of the ribbon, $\hat{K}_1$ remains approximately zero throughout.

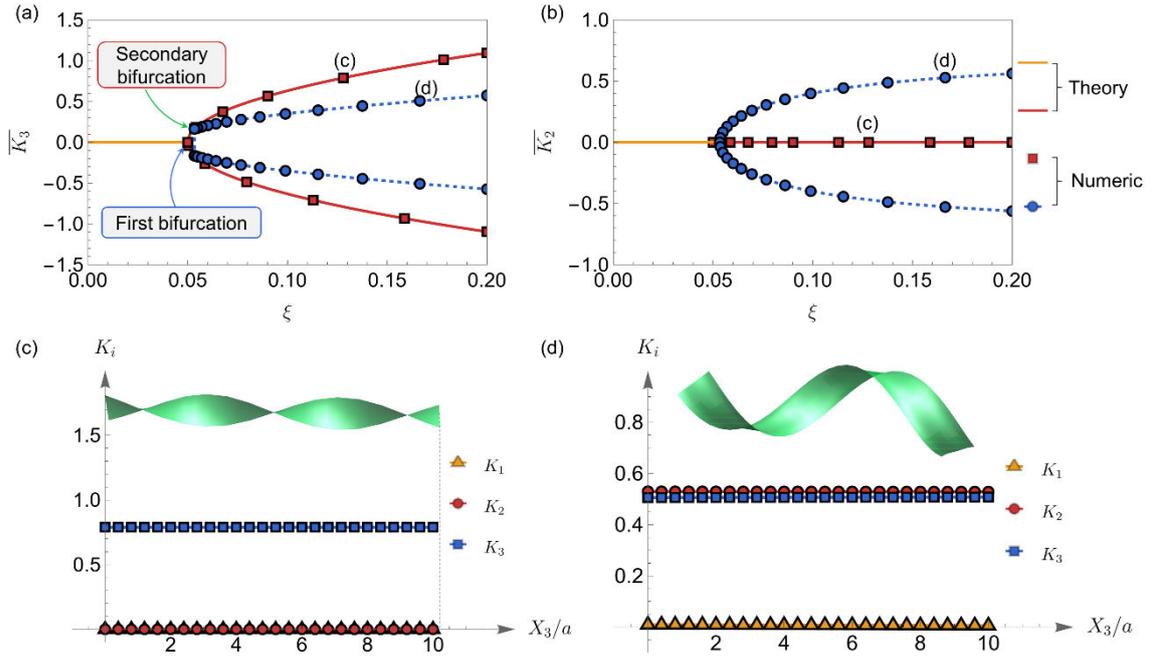

*Fig. 10 Bifurcation diagrams for the ribbon with different growth strains in the longitudinal direction that exhibits a gradient along the $X_2$-axis. (a) Variation of the average twisting curvature $\bar{K}_3$ to the growth strain. (b) Variation of the average curvature $\bar{K}_2$ to the growth strain. (c) Distribution of curvature along the longitudinal direction for twisting ribbons. (d) Distribution of curvature along the longitudinal direction for helical ribbons.*

The morphological evolution of ribbons with different thickness-to-width ratios under differential growth is shown in Fig. 11. As shown in Fig. 11(a), with the gradual increase of $\xi$, the ribbon with $\eta = 0.05$ transitions from a planar to a twisted configuration at $\xi > 0.125$. When $\xi$ exceeds approximately 0.013, a secondary bifurcation occurs, driving the transition from a twisting to a helical configuration. In the helical regime, the normalized out-of-plane displacement $u_1/a$ along the longitudinal direction remains consistent, while its distribution across the width exhibits notable variation. Initially, the displacement is smaller near the edge region (approximately $0.4 < X_2/a < 0.5$) and larger near the center. As $\xi$ increases, due to changes in the intrinsic curvature and torsion, the displacement near the edges gradually decreases, while the value in the central region evolves from positive to negative. For $\eta = 0.1$ (see Fig. 11(b)), two similar morphological transitions are observed around $\xi = 0.052$ and $0.054$, respectively. Moreover, the variation in cross-sectional displacement becomes more pronounced compared to the case with smaller $\eta$.

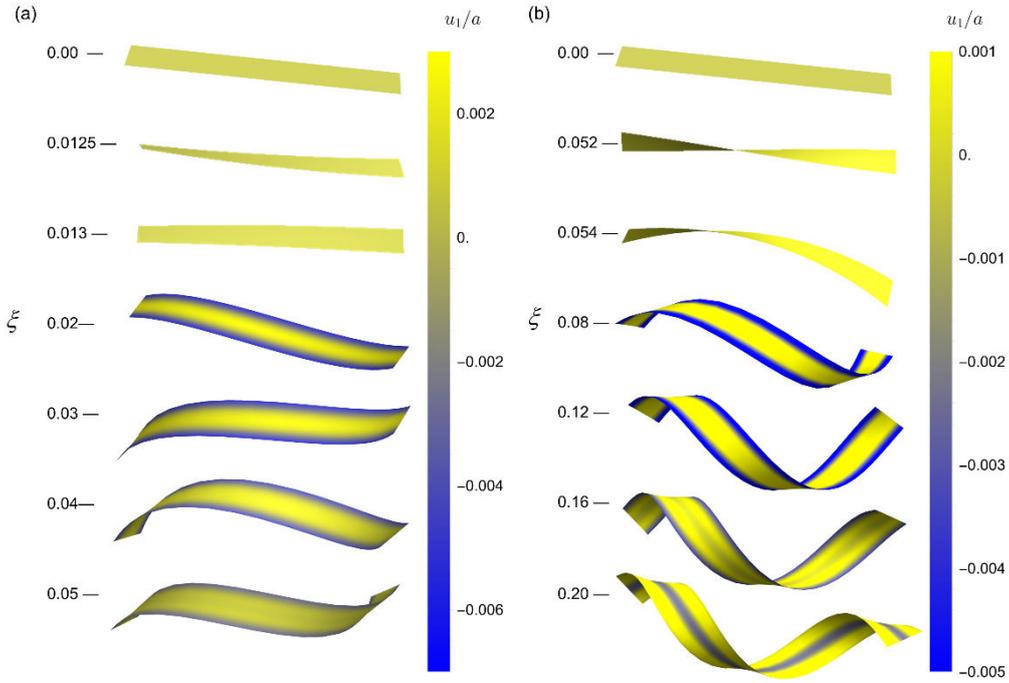

Fig. 11 Morphological evolution of the helical ribbon: (a) $\eta = 0.05$ and (b) $\eta = 0.1$.

The morphological phase diagram of the ribbon under longitudinal growth strain exhibiting a gradient along the $X_2$- axis is shown in Fig. 12. The blue and red lines indicate the critical values corresponding to the primary and secondary bifurcations, respectively. It is observed that as $\eta$ increases, the critical growth strains for both bifurcations increase nonlinearly. For a given thickness-to-width ratio, as $\xi$ increases, the ribbon undergoes a sequential morphological transition—from a planar configuration to a twisted state and eventually to a helical shape.

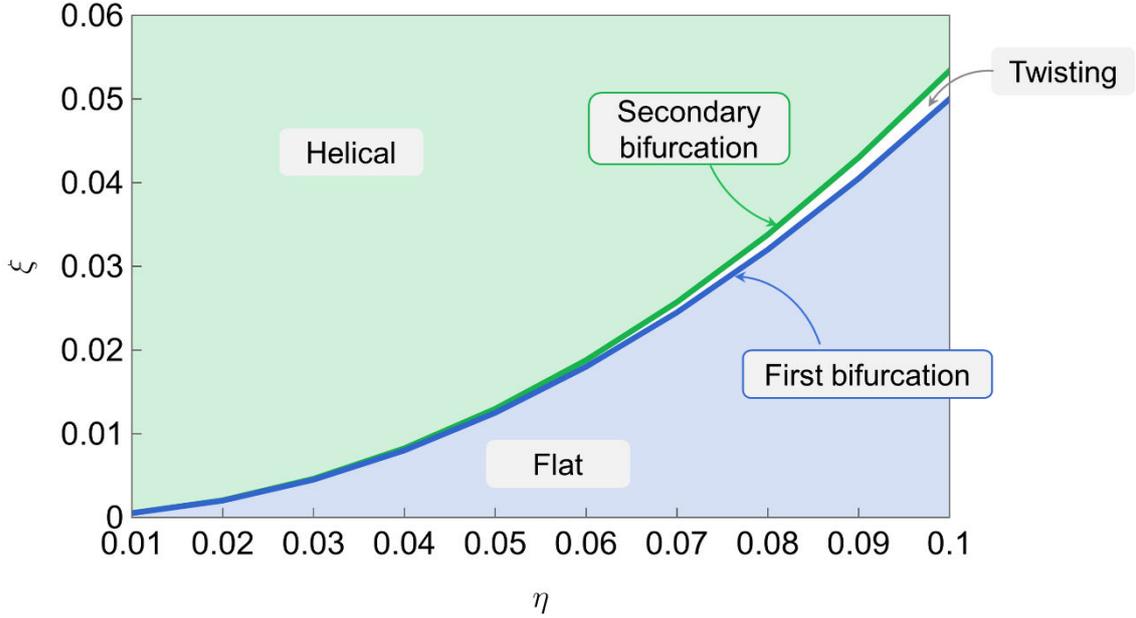

*Fig. 12 Phase diagram of the ribbon with different growth strains in the longitudinal direction that exhibits a gradient along the $X_2$ - axis*

## 4 Conclusion

Leaves and petals of plants typically exhibit slender, ribbon-like geometries characterized by a thickness that is significantly smaller than their width and a width that is significantly smaller than their length. In this work, we developed a 1D morphoelastic ribbon model based on morphoelasticity theory. Starting from three-dimensional nonlinear elasticity, the deformed configuration is described in macroscopic curvature along the centerline and local displacements within the cross-section. By introducing the thickness-to-width ratio as a small parameter, we first reduce the three-dimensional problem to a two-dimensional shell model, where the strain distribution within the cross-section can asymptotically capture the nonlinear effects induced by finite rotations, with accuracy up to second order in the thickness-to-width ratio. We then employed a second asymptotic reduction using the slenderness parameter (width-to-length ratio), yielding a complete 1D model. Within this framework, we established a connection between the local growth strain field and the global stretching of the centerline, as well as the intrinsic curvatures induced by differential growth. From a modeling perspective, the dimension reduction approach is general and widely applicable to slender morphoelastic structures.

Building on this one-dimensional model, we systematically analyzed a range of growth-induced morphologies, including saddle-like bending, twisting, and helical configurations, driven by growth strain gradients along different directions. Analytical solutions were derived for intrinsic curvature induced by differential growth across the thickness, highlighting the dependence of curvature on both the growth field and cross-sectional geometry. We further identified twisting configurations driven by growth gradients across the width and derived the corresponding critical growth strain for the onset of twisting. The emergence of helical shapes is interpreted as a secondary bifurcation arising with increasing growth strain. The simplicity of the 1D model enabled efficient numerical implementation, allowing us to capture the evolution of intrinsic curvature and torsion throughout the helical transition. Finally, we construct a phase diagram that maps the morphological transitions as functions of growth strain and the thickness-to-width ratio.

The framework established in this study offers a general analytical approach for modeling growth-induced deformations in slender ribbons. Our morphoelastic ribbon model successfully decouples the competing effects of differential growth and geometric nonlinearities in governing the 3D morphological transitions—curvature generation, torsion development, and helical bifurcations—thereby establishing a unified mechanics-genesis relationship in slender plant organs. The proposed model also holds promise for the bioinspired design of soft robots.

## CRediT authorship contribution statement

**Hao Liu:** Writing – original draft, Visualization, Software, Methodology, Investigation, Formal analysis, Data curation, Conceptualization. **Mingwu Li**: Writing – review & editing, Software, Formal analysis. **Dabiao Liu:** Writing – review & editing, Supervision, Methodology, Investigation, Formal analysis, Conceptualization

## Declaration of competing interest

The authors declare that they have no known competing financial interests or personal relationships that could have appeared to influence the work reported in this paper.

## Acknowledgments

This work is financially supported by the National Natural Science Foundation of China (Grant No. 12272146), the Fundamental Research Funds for the Central Universities (Grant No. 2024BRA009 and YCJJ20252105), and the National Ten Thousand Talent Program for Young Top-notch Talents.

## Appendix A. Detailed derivation of the reduced shell model

By differentiating the deformation $\chi$ with respect to $\mathbf{X}$, we have the deformation gradient

$$\mathbf{F} = F_{ij}\mathbf{d}_i \otimes \mathbf{E}_j = \begin{bmatrix} 1+U_{1,X_1} & U_{1,X_2} & U_{1,X_3}+U_3\kappa_2-(X_2+U_2)\kappa_3 \\ U_{2,X_1} & 1+U_{2,X_2} & U_{2,X_3}-U_3\kappa_1+(X_1+U_1)\kappa_3 \\ U_{3,X_1} & U_{3,X_2} & 1+\varepsilon+U_{3,X_3}+(X_2+U_2)\kappa_1-(X_1+U_1)\kappa_2 \end{bmatrix}. \quad (71)$$

where $U_{i,X_j} = \frac{\partial U_i}{\partial X_j}$. Based on the multiplicative decomposition proposed by Rodriguez et al. (1994), we have the elastic deformation $\mathbf{A} = \mathbf{F} \cdot \mathbf{G}^{-1}$. Then, the fully nonlinear strain can be computed using $\mathbf{E} = \frac{1}{2}(\mathbf{A}^T \cdot \mathbf{A} - \mathbf{I})$. They are

$$\begin{aligned} E_{11} = \frac{1}{2}\Big(&\big(g_{31}(U_{1,X_3}+\kappa_2 U_3-\kappa_3(U_2+X_2))+g_{21}U_{1,X_2}+(g_{11}-1)(\zeta_{,X_1}+1)\big)^2 \\ &+\big(g_{31}(\kappa_3(U_1+X_1)+U_{2,X_3}-\kappa_1 U_3)+g_{21}(U_{2,X_2}+1)+(g_{11}-1)U_{2,X_1}\big)^2 \\ &+\big(g_{31}(-\kappa_2(U_1+X_1)+U_{3,X_3}+\kappa_1(U_2+X_2)+\varepsilon+1)+g_{21}U_{3,X_2}+(g_{11}-1)U_{3,X_1}\big)^2 - 1\Big) \end{aligned} \quad (72)$$

$$\begin{aligned} E_{12} = E_{21} = \frac{1}{2}\Big(&\big(-g_{31}(\kappa_3(U_1+X_1)+U_{2,X_3}-\kappa_1 U_3)-g_{21}(U_{2,X_2}+1)-(g_{11}-1)U_{2,X_1}\big) \\ &\big(-g_{32}(-U_3\kappa_1+\kappa_3(X_1+U_1)+U_{2,X_3})+(1-g_{22})(1+U_{2,X_2})-g_{12}U_{2,X_1}\big) \\ &+\big(-g_{31}(1+(U_2+X_2)\kappa_1-\kappa_2(X_1+U_1)+\varepsilon+U_{3,X_3})-g_{21}U_{3,X_2}+(1-g_{11})U_{3,X_1}\big) \\ &\big(-g_{32}(1+(U_2+X_2)\kappa_1-\kappa_2(X_1+U_1)+\varepsilon+U_{3,X_3})+(1-g_{22})U_{3,X_2}-g_{12}U_{3,X_1}\big) \\ &+\big(-g_{31}(U_3\kappa_2-(U_2+X_2)\kappa_3+U_{1,X_3})-g_{21}U_{1,X_2}+(1-g_{11})(1+\zeta_{,X_1})\big) \\ &\big(-g_{32}(U_3\kappa_2-(U_2+X_2)\kappa_3+U_{1,X_3})+(1-g_{22})U_{1,X_2}-g_{12}(1+\zeta_{,X_1})\big)\Big) \end{aligned} \quad (73)$$

$$E_{22} = \frac{1}{2}\Big(\big(g_{32}(U_{1,X_3} + \kappa_2 U_3 - \kappa_3(U_2 + X_2)) + (g_{22}-1)U_{1,X_2} + g_{12}(\zeta_{,X_1}+1)\big)^2$$
$$+ \big(g_{32}(\kappa_3(U_1+X_1) + U_{2,X_3} - \kappa_1 U_3) + (g_{22}-1)(U_{2,X_2}+1) + g_{12}U_{2,X_1}\big)^2 \quad (74)$$
$$+ \big(g_{32}(-\kappa_2(U_1+X_1) + U_{3,X_3} + \kappa_1(U_2+X_2) + \varepsilon + 1) + (g_{22}-1)U_{3,X_2} + g_{12}U_{3,X_1}\big)^2 - 1\Big)$$

$$E_{13} = E_{31} = \frac{1}{2}\Big(\big(-g_{31}(-U_3\kappa_1 + \kappa_3(X_1+U_1) + U_{2,X_3}) - g_{21}(1+U_{2,X_2}) + (1-g_{11})U_{2,X_1}\big)$$
$$\big((1-g_{33})(-U_3\kappa_1 + \kappa_3(X_1+U_1) + U_{2,X_3}) - g_{23}(1+U_{2,X_2}) - g_{13}U_{2,X_1}\big)$$
$$+ \big(-g_{31}(1+(U_2+X_2)\kappa_1 - \kappa_2(X_1+U_1) + \varepsilon + U_{3,X_3}) - g_{21}U_{3,X_2} + (1-g_{11})U_{3,X_1}\big) \quad (75)$$
$$\big((1-g_{33})(1+(U_2+X_2)\kappa_1 - \kappa_2(X_1+U_1) + \varepsilon + U_{3,X_3}) - g_{23}U_{3,X_2} - g_{13}U_{3,X_1}\big)$$
$$+ \big(-g_{31}(U_3\kappa_2 - (U_2+X_2)\kappa_3 + U_{1,X_3}) - g_{21}U_{1,X_2} + (1-g_{11})(1+\zeta_{,X_1})\big)$$
$$\big((1-g_{33})(U_3\kappa_2 - (U_2+X_2)\kappa_3 + U_{1,X_3}) - g_{23}U_{1,X_2} - g_{13}(1+\zeta_{,X_1})\big)\Big)$$

$$E_{32} = E_{23} = \frac{1}{2}\Big(\big(-g_{32}(-U_3\kappa_1 + \kappa_3(X_1+U_1) + U_{2,X_3}) + (1-g_{22})(1+U_{2,X_2}) - g_{12}U_{2,X_1}\big)$$
$$\big((1-g_{33})(-U_3\kappa_1 + \kappa_3(X_1+U_1) + U_{2,X_3}) - g_{23}(1+U_{2,X_2}) - g_{13}U_{2,X_1}\big)$$
$$+ \big(-g_{32}(1+(U_2+X_2)\kappa_1 - \kappa_2(X_1+U_1) + \varepsilon + U_{3,X_3}) + (1-g_{22})U_{3,X_2} - g_{12}U_{3,X_1}\big) \quad (76)$$
$$\big((1-g_{33})(1+(U_2+X_2)\kappa_1 - \kappa_2(X_1+U_1) + \varepsilon + U_{3,X_3}) - g_{23}U_{3,X_2} - g_{13}U_{3,X_1}\big)$$
$$+ \big(-g_{32}(U_3\kappa_2 - (U_2+X_2)\kappa_3 + U_{1,X_3}) + (1-g_{22})U_{1,X_2} - g_{12}(1+\zeta_{,X_1})\big)$$
$$\big((1-g_{33})(U_3\kappa_2 - (U_2+X_2)\kappa_3 + U_{1,X_3}) - g_{23}U_{1,X_2} - g_{13}(1+\zeta_{,X_1})\big)\Big)$$

$$E_{33} = \frac{1}{2}\Big(\big((g_{33}-1)(U_{1,X_3} + \kappa_2 U_3 - \kappa_3(U_2+X_2)) + g_{23}U_{1,X_2} + g_{13}(\zeta_{,X_1}+1)\big)^2$$
$$+ \big((g_{33}-1)(\kappa_3(U_1+X_1) + U_{2,X_3} - \kappa_1 U_3) + g_{23}(U_{2,X_2}+1) + g_{13}U_{2,X_1}\big)^2 \quad (77)$$
$$+ \big((g_{33}-1)(-\kappa_2(U_1+X_1) + U_{3,X_3} + \kappa_1(U_2+X_2) + \varepsilon + 1) + g_{23}U_{3,X_2} + g_{13}U_{3,X_1}\big)^2 - 1\Big)$$

Following Audoly and Neukirch (2021), we begin with the theory of elastic rods to determine the scaling of axial strain $\varepsilon$ and curvatures $\kappa_i$. For a rectangular cross-section ($a \times h$), the stretching modulus is $Y_e a h$, the bending moduli scale as $Y_e a^3 h$ and $Y_e a h^3$, and the torsional modulus is of the order $Y_e a h^3$. Assuming that the contributions of axial strain and curvature to the elastic energy are of the same order

$$Y_e h a \varepsilon^2 \sim Y_e a^3 h \kappa_1^2 \sim Y_e a h^3 \kappa_2^2 \sim Y_e a h^3 \kappa_3^2. \quad (78)$$

We obtain that $\kappa_2 \sim \kappa_3$, $\kappa_1 \sim \eta \kappa_2$, $\varepsilon \sim h\kappa_2$.

For a thin shell structure, we adopt the von Kármán assumptions, namely that the in-plane elastic strains on the mid-surface are minor while the transverse deflections are large. Accordingly, we retain only the gradients of the out-of-plane displacement $u_1(X_2, X_3)$ in the nonlinear part of the elastic strain tensor so that their orders of magnitude satisfy: $U_2 \sim U_3 \sim u_1^2/a \sim \eta^2 a$. The ribbon is nearly incompressible with Poisson's ratio $v = 0.5$. Therefore, the volume change should be zero, i.e., $\det(\mathbf{F}) = 1$. This implies $\varepsilon \sim \eta^2$, $\kappa_1 \sim \eta^2/a$, $\kappa_2 \sim \kappa_3 \sim \eta/a$ and $\zeta \sim \eta^3 a$.

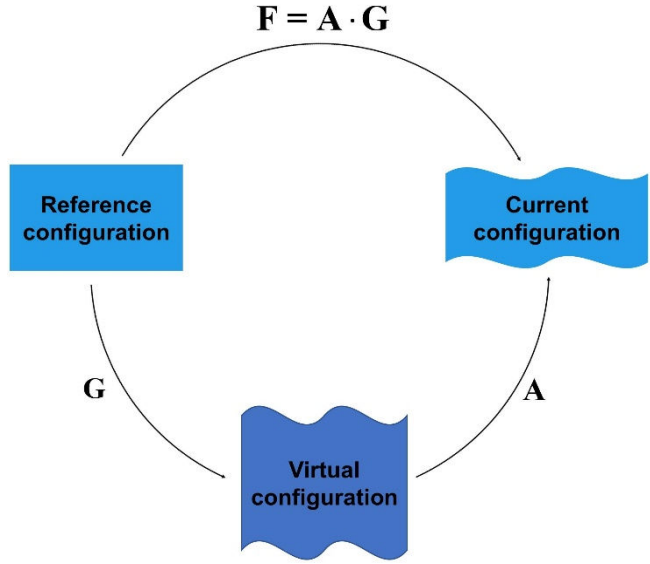

**Fig A1.** Sketch of multiplicative decomposition of deformation of a ribbon.

We consider compatible growth where the virtual configuration coincides with the current one in Fig. A1. In this case, $\mathbf{g}$ should be identical to the gradient of some displacement field with zero deflection $u_1 = 0$. Then one has $g_{\alpha\beta} \sim \eta^2$, $g_{\alpha 1} \sim \eta$, $g_{1\beta} \sim \eta^3$ and $g_{11} \sim \eta^2$. Based on the above scaling analysis, we obtain the simplified expression of the weakly nonlinear strain used in the main text (Eq. (14))

## Appendix B. Variational derivatives of the ribbon model

Four scalar Lagrange multipliers $(f_1, f_2, f_3, f_4)$ are introduced to deal with kinematic constraints of the displacement in Eqs. (16) and (17). By Eq (25), we have the augmented total potential energy expressed as

$$\mathcal{L}=\int_0^{l_0}\int_{-a/2}^{a/2}\int_{-h/2}^{h/2}\left(\Psi\left(u_1,u_2,u_3;\varepsilon,\kappa_1,\kappa_2,\kappa_3;\mathbf{g}\right)+f_1u_1+f_2u_2+f_3u_3+f_4X_2u_1\right)\mathrm{d}X_1\mathrm{d}X_2\mathrm{d}X_3. \quad(79)$$

Next, we expand the macroscopic and Lagrange multipliers as a series expansion in powers of $\gamma$,

$$\begin{aligned}
u_1 &= u_1^{(0)} + \gamma u_1^{(1)} + \gamma^2 u_1^{(2)} + \dots \\
u_2 &= u_2^{(0)} + \gamma u_2^{(1)} + \gamma^2 u_2^{(2)} + \dots \\
u_3 &= u_3^{(0)} + \gamma u_3^{(1)} + \gamma^2 u_3^{(2)} + \dots \\
p_i &= p_i^{(0)} + \gamma p_i^{(1)} + \gamma^2 p_i^{(2)} + \dots
\end{aligned} \quad(80)$$

where $p_i=\{f_1,f_2,f_3,f_4\}$.

Here, we only consider the solution of the displacement to order of $\left(u_1^{(0)},u_2^{(0)},u_3^{(0)}\right)$. We now perform a variational derivation of the hyperelastic ribbon by perturbing the displacements $(u_1,u_2,u_3)\to(u_1+\delta u_1,u_2+\delta u_2,u_3+\delta u_3)$. The first variation of the total potential energy reads

$$\delta\mathcal{L}=\int_0^{l_0}\int_{-a/2}^{a/2}\int_{-h/2}^{h/2}\left(\Psi\left(\delta u_1,\delta u_2,\delta u_3;\varepsilon,\kappa_1,\kappa_2,\kappa_3;\mathbf{g}\right)+f_1\delta u_1+f_2\delta u_2+f_3\delta u_3+f_4X_2\delta u_1\right)\mathrm{d}X_1\mathrm{d}X_2\mathrm{d}X_3. \quad(81)$$

Submitting the right Cauchy-Green deformation tensor and curvature into Eq (81), and requiring $\delta\mathcal{L}=0$ for all admissible displacements $(\delta u_1,\delta u_2,\delta u_3)$, we obtain the Euler–Lagrange equations

$$\frac{Y_e h}{3}\left(g_{23}+g_{32}-\kappa_3 u_1 - u_{3,X_2}+\kappa_3 X_2 u_{1,X_2}\right)_{,X_2}+f_3=0, \quad(82)$$

$$\frac{Y_e h}{3}\left(2g_{33}+4g_{22}-2\varepsilon-2X_2\kappa_1+2\kappa_2 u_1-X_2^2\kappa_3^2-4u_{2,X_2}-2u_{1,X_2}^2\right)_{,X_2}+f_2=0, \quad(83)$$

$$\frac{Y_e h^3}{18}\left[\kappa_2+2u_{1,X_2X_2}-2g_{21,X_2}\right]_{,X_2X_2}+Y_e h\kappa_2^2 u_1+Y_e h\kappa_2\left(g_{33}-\varepsilon-\kappa_1 X_2-\frac{\kappa_3^2 X_2^2}{2}+\frac{f_1+X_2f_4}{Y_e h\kappa_2}\right)=0. \quad(84)$$

and the boundary conditions

$$\forall X_3\in[0,l_0],\ X_2=\pm a/2:\quad g_{23}+g_{32}-\kappa_3 u_1-u_{3,X_2}+\kappa_3 X_2 u_{1,X_2}=0, \quad(85)$$

$$\forall X_3\in[0,l_0],\ X_2=\pm a/2:\quad 2g_{33}+4g_{22}-2\varepsilon-2X_2\kappa_1+2\kappa_2 u_1-X_2^2\kappa_3^2-4u_{2,X_2}-2u_{1,X_2}^2=0, \quad(86)$$

$$\forall X_3\in[0,l_0],\ X_2=\pm a/2:\quad \kappa_2+2u_{1,X_2X_2}-2g_{21,X_2}=0, \quad(87)$$

$$\forall X_3\in[0,l_0],\ X_2=\pm a/2:\quad \left(\kappa_2+2u_{1,X_2X_2}-2g_{21,X_2}\right)_{,X_2}=0. \quad(88)$$

## Data availability

Data will be made available on request.